\documentclass[12pt,preprint]{aastex}

\shorttitle{A Catalog of BALQSOs from the SDSS DR3}
\shortauthors{Trump et~al.}

\begin{document}

\title{A Catalog of Broad Absorption Line Quasars from the Sloan Digital Sky Survey Third Data Release}

\author{
Jonathan R. Trump,\altaffilmark{\ref{PSU}}$^,$\altaffilmark{\ref{Arizona}}
Patrick B. Hall,\altaffilmark{\ref{Princeton}}$^,$\altaffilmark{\ref{York}}
Timothy A. Reichard,\altaffilmark{\ref{JHU}}
Gordon T. Richards,\altaffilmark{\ref{Princeton}}$^,$\altaffilmark{\ref{JHU}}
Donald P. Schneider,\altaffilmark{\ref{PSU}}
Daniel E. Vanden Berk,\altaffilmark{\ref{PSU}}
Gillian R. Knapp,\altaffilmark{\ref{Princeton}}
Scott F. Anderson,\altaffilmark{\ref{UWash}}
Xiaohui Fan,\altaffilmark{\ref{Arizona}}
J. Brinkman,\altaffilmark{\ref{APO}}
S.~J. Kleinman,\altaffilmark{\ref{Subaru}}
and Atsuko Nitta\altaffilmark{\ref{Subaru}}
}

\altaffiltext{1}{
  Department of Astronomy and Astrophysics, The Pennsylvania State
  University, University Park, PA 16802.
\label{PSU}}

\altaffiltext{2}{
  Steward Observatory, University of Arizona, 933 North Cherry Avenue,
  Tucson, AZ 85721
\label{Arizona}}

\altaffiltext{3}{
  Princeton University Observatory, Princeton, NJ 08544.
\label{Princeton}}

\altaffiltext{4}{
  Department of Physics \& Astronomy, York University, 4700 Keele
  Street, Toronto, Ontario, M3J 1P3, Canada
\label{York}}

\altaffiltext{5}{
  The Johns Hopkins University, 3400 North Charles Street, Baltimore,
  MD 21218-2686
\label{JHU}}

\altaffiltext{6}{
  Department of Astronomy, University of Washington, Box 351580,
  Seattle, WA 98195
\label{UWash}}

\altaffiltext{7}{
  Apache Point Observatory, P.O. Box 59, Sunspot, NM 88349-0059.
\label{APO}}

\altaffiltext{8}{
  Subaru Telescope, 650 N A'ohoku Pl., Hilo, HI 96720
\label{Subaru}}

%\altaffiltext{6}{
%   Department of Physics \& Astronomy, The University of Pittsburgh,
%   3941 O'Hara Street, Pittsburgh, PA, 15260
%\label{Pitt}}

\newcommand{\CIV}{\hbox{{\rm C}\kern 0.1em{\sc iv}}}
\newcommand{\MgI}{\hbox{{\rm Mg}\kern 0.1em{\sc i}}}
\newcommand{\MgII}{\hbox{{\rm Mg}\kern 0.1em{\sc ii}}}
\newcommand{\FeII}{\hbox{{\rm Fe}\kern 0.1em{\sc ii}}}
\newcommand{\FeIII}{\hbox{{\rm Fe}\kern 0.1em{\sc iii}}}
\newcommand{\NV}{\hbox{{\rm N}\kern 0.1em{\sc v}}}
\newcommand{\SiIV}{\hbox{{\rm Si}\kern 0.1em{\sc iv}}}
\newcommand{\OIV}{\hbox{{\rm O}\kern 0.1em{\sc iv}]}}
\newcommand{\AlIII}{\hbox{{\rm Al}\kern 0.1em{\sc iii}}}
\newcommand{\CIII}{\hbox{{\rm C}\kern 0.1em{\sc iii}]}}
\newcommand{\NeV}{\hbox{{\rm [Ne}\kern 0.1em{\sc v}]}}
\newcommand{\OII}{\hbox{{\rm [O}\kern 0.1em{\sc ii}]}}
\newcommand{\NeIII}{\hbox{{\rm [Ne}\kern 0.1em{\sc iii}]}}
\newcommand{\Lya}{\hbox{{\rm Ly}\kern 0.1em$\alpha$}}
\newcommand{\Lyb}{\hbox{{\rm Ly}\kern 0.1em$\beta$}}
\newcommand{\Hd}{\hbox{{\rm H}\kern 0.1em$\delta$}}
\newcommand{\kms}{\hbox{km~s$^{-1}$}}

\begin{abstract}

%%4386 CIV + 457 MgII - 98 bothCIV&MgII + 39 z>4.38 = 4784 unique BALs

We present a total of 4784 unique broad absorption line quasars from
the Sloan Digital Sky Survey Third Data Release.  An automated
algorithm was used to match a continuum to each quasar and to identify
regions of flux at least 10\% below the continuum over a velocity
range of at least $1000~\kms$ in the $\CIV$ and $\MgII$ absorption
regions.  The model continuum was selected as the best-fit match from
a set of template quasar spectra binned in luminosity, emission line
width, and redshift, with the power-law spectral index and amount of
dust reddening as additional free parameters.  We characterize our
sample through the traditional ``balnicity" index and a revised
absorption index, as well as through parameters such as the width,
outflow velocity, fractional depth and number of troughs.  From a
sample of 16883 quasars at $1.7\le z\le 4.38$, we identify 4386
($26.0\%$) quasars with broad $\CIV$ absorption, of which 1756
($10.4\%$) satisfy traditional selection criteria.  From a sample of
34973 quasars at $0.5\le z\le 2.15$, we identify 457 ($1.31\%$)
quasars with broad $\MgII$ absorption, 191 ($0.55\%$) of which satisfy
traditional selection criteria.  We also provide a supplementary list
of 39 visually identified $z>4.38$ quasars with broad $\CIV$
absorption.  We find that broad absorption line quasars may have
broader emission lines on average than other quasars.

\end{abstract}


\section{Introduction}

The nature of intrinsic absorption in quasars and other active
galactic nuclei (AGN) has important implications for physical models
of the quasar ``central engine.''  The subclass of broad absorption
line quasars (BALQSOs) is thought to account for between 10\% and 30\%
of quasars.  The exact fraction of BALQSOs is nontrivial to determine
because of differential selection effects between BALQSO and
non-BALQSOs.  Hewett \& Foltz (2003) report a corrected BALQSO
fraction of $22 \pm 4 \%$ for their sample of 42 bright ($B_J < 19$)
quasars, Tolea, Krolik \& Tsvetanov (2002) estimate a fraction of
$\sim15\%$ from their sample of 116 quasars, and Reichard et
al. (2003b) estimate a corrected fraction of $15.9 \pm 1.4 \%$ for a
sample of 224 quasars with $i\lesssim20$.

Broad absorption troughs are caused by gas outflowing at high
velocities from quasars.  It has been suggested that some or all
BALQSOs may be a unique class of quasars (Surdej \& Hutsemekers 1987),
perhaps in a different stage of their life cycle (Hazard et~al. 1984,
Boroson \& Meyers 1992, Becker et~al. 2000).  Other authors suggest
the differences from ``standard'' quasars arise because BALQSOs are
observed at a different orientation (Weymann et~al. 1991; Ogle
et~al. 1999; Schmidt \& Hines 1999; Hall et~al. 2002).  BALQSOs
generally have UV to soft X-ray flux ratios 10-30 times smaller than
unabsorbed quasars (Brandt, Laor, \& Wills 2000), though if BALQSOs
are corrected for intrinsic absorption, their UV to soft X-ray flux
ratio is typical of normal quasars (Gallagher et~al. 2002).

Three sources of material can cause absorption in the spectrum of a
quasar: (1) ``intrinsic,'' produced by processes related to the AGN
itself, (2) ``host,'' produced by material in the quasar's host galaxy
unrelated to the AGN, and (3) ``intervening,'' absorbing material
along the line of sight but not physically related to the AGN or host
galaxy.  Broader and higher-velocity absorption troughs are more
likely to be intrinsic, since only the AGN can feasibly accelerate
matter to the velocities and velocity widths observed in BALQSOs (both
often $\ge 10000~\kms$).
%Though we avoid using the term herein,
%absorption occurring within $5000~\kms$ of the quasar redshift is
%often termed ``associated absorption'' (Foltz et~al. 1986).

BALQSOs are classified into three subcategories based upon the
material producing the BAL troughs.  High-ionization BALQSOs (HiBALs)
contain strong, broad absorption troughs shortward of high-ionization
emission lines and are typically identified through the presence of
$\CIV$ absorption troughs.  Low-ionization BALQSOs (LoBALs) contain
HiBAL features but also have absorption from low-ionization lines such
as $\MgII$.  LoBALs with excited-state $\FeII$ or $\FeIII$ absorption
are called FeLoBALs.

The first large sample of BALQSOs was analyzed by Weymann et~al.
(1991), who defined BALQSOs as quasars exhibiting $\CIV$ absorption
troughs broader than 2000 $\kms$.  This minimum trough width was
defined primarily to avoid contamination by noise, but also to reject
multiple overlapping intervening systems.  We call these objects
``traditional'' BALQSOs; they are fully defined in \S 4.1.  Reichard
et~al. (2003a) expanded the number of known traditional BALQSOs with a
catalog of 224 BALQSOs selected from the Sloan Digital Sky Survey
(SDSS; York et~al. 2000) Early Data Release (EDR; Stoughton et
al. 2002a) quasar catalog (Schneider et~al. 2002).  (See Menou et
al. (2001) and Tolea, Krolik \& Tsvetanov (2002) for earlier SDSS work
on BALQSOs.)

In this paper we present a catalog of 4784 BALQSOs from the SDSS Third
Data Release (DR3; Abazajian et~al. 2005).  Our catalog represents a
substantial increase in the number of known BALQSOs primarily because
the SDSS DR3 is more than ten times larger than the SDSS EDR.  We
retain the traditional BALQSO definition as all quasars with
``balnicity index" ${\rm BI} > 0$ (Weymann et~al. 1991, also defined
in \S 4.1), but also extend our analysis of the intrinsic absorption
lines in quasars to include absorption troughs broader than 1000
$\kms$ regardless of their velocity shift from the quasar redshift.
Our full definition of BALQSOs is given in \S 4.3.  Our automated
algorithms identify HiBALs via $\CIV$ from $1.7 \le z \le 4.38$ and
LoBALs via $\MgII$ from $0.5 \le z \le 2.15$.  These redshift
constraints are set by the wavelength coverage of the SDSS spectra
(3800-9200 \AA).  FeLoBAL objects can have strong absorption
throughout their spectra because there are lots of Fe lines in several
ionization states.  FeLoBALs are easily identified through visual
inspection of HiBALs and LoBALs, but are very difficult to fit with a
continuum in the automated methods.

The parent sample for our catalog is discussed in \S 2.  In \S 3 we
describe the automated BALQSO identification scheme and motivate the
expansion of the BALQSO definition.  We define the metrics used to
describe both the traditional BALQSOs and the BALQSOs of the catalog
in \S 4.  The different objects included in the catalog are discussed
in \S 5.  We suggest some of the implications of our statistical
sample of BALQSOs in \S 6, though a more complete analysis of the
sample will be presented by Hall et~al. (2006).  Throughout this
paper, we adopt a cosmology consistent with WMAP results of $h=0.70$,
$\Omega_M=0.3$, $\Omega_{\Lambda}=0.7$ (Spergel et~al. 2003).

\section{Observations: The Sloan Digital Sky Survey}

The parent sample for the BALQSO catalog is the SDSS DR3 quasar
catalog (Schneider et~al. 2005).  The SDSS is a wide-field survey
operated by the Astrophysical Research Consortium on a 2.5 m telescope
(Gunn et~al. 2006) at the Apache Point Observatory, New Mexico.  All
images are taken in the five bands \emph{u}, \emph{g}, \emph{r},
\emph{i}, \emph{z} (Fukugita et~al. 1996) by the CCD camera (Gunn et
al. 1998).  SDSS magnitudes are expressed in the asinh magnitude
system (Lupton, Gunn, \& Szalay 1999).  Calibration of the photometry
is described by Hogg et~al. (2001), Smith et~al. (2002), and Tucker et
al. (2006), photometric processing by Lupton et al. (2001), and the
photometric quality assessement by Ivezi\'c et al. (2004).  Quasar
candidates are selected for spectroscopic targeting based on the
selection algorithm of Richards et~al. (2002).  Spectroscopy of all
objects in the DR3 was acquired via 826 $3^\circ$ diameter plates with
640 drilled fibers per plate, with the plates tiled onto the sky using
the algorithm of Blanton et~al. (2003).  The spectroscopic pipeline is
discussed by Stoughton et~al. (2002a).  The resolution of the SDSS
spectra is $1800-2100$ over most of the wavelength coverage, 3800-9200
\AA.
%The accuracy of the spectrophotometry in the DR3 is increased over
%previous SDSS releases due to the improved comparison of fiber
%magnitudes to photometry.  
%Spectra are not corrected for Galactic reddening, which has a median
%value $E(B-V)=0.034$ over the sample (Abazajian et~al. 2004).

The SDSS DR3 quasar catalog contains all 46420 spectroscopically
identified quasars in the DR3 with $M_{\rm i} < -22$.  We correct all
spectra for the Galactic extinction given in the maps of Schlegel,
Finkbeiner, \& Davis (1998) using the empirical selective extinction
function of Cardelli, Clayton, \& Mathis (1989).  We boxcar smooth all
spectra by $\sim 3$ pixels (roughly the resolution) to reduce noise
and prevent the use of unresolved (spurious) features.  The SDSS
spectroscopic pipeline automatically interpolates over bad pixels,
although on either side of an interpolated region there are typically
a few pixels with a very low signal-to-noise ratio.  To correct this
we extend both sides of all pipeline-interpolated regions by three
pixels and ignore the entire extended interpolated regions.  Our
analyses are limited to SDSS spectra in which the both the region up
to 29000 $\kms$ shortward of the $\MgII$ or $\CIV$ emission line and
the normalization region ($2910\pm10$ \AA\ for $\MgII$ and $1710\pm10$
\AA\ for $\CIV$) are redshifted into the observed spectral range.
Thus we searched for $\MgII$ BALQSOs among 34973 objects with $0.5 \le
z \le 2.15$ and $\CIV$ BALQSOs among 16883 objects with $1.7 \le z \le
4.38$.

\section{Construction of the Catalog}

The need for objective selection and quantification of BALQSOs and
traditional BALQSOs in our large data set necessitated the use of an
automatic selection algorithm.  Our basic approach was that of
Reichard et~al. (2003a): a continuum fit was made to each quasar
spectrum and absorption features were identified as dips in the
spectrum below the continuum.  But whereas Reichard et~al. (2003a)
used a single template spectrum to fit each sample continuum, we
employ a best-fit template from a set of template spectra binned in
line width, luminosity, and redshift.  We also extend the fitting
ranges from those of Reichard et~al. (2003a) to better account for
line shapes by including emission line profiles redward of the center
wavelength (where absorption should not play a significant role).  We
also independently scale the $\CIV$ and $\MgII$ emission line regions
of the best-fit template.  The template spectra are described in
detail in \S 3.1 and the fitting process in \S 3.2.

\subsection{Template Spectra}

Most quasars have broadly similar ultraviolet/optical spectra (e.g.,
Richards et~al. 2001, Vanden Berk et~al. 2001), but there are several
basic trends which lead to considerable variation among quasar spectra
(Marziani et~al. 2003).  One of the most prominent is the Baldwin
(1977) Effect, the well-known anticorrelation between the luminosity
of a quasar and the equivalent widths of its emission lines (see also
Osmer, Porter, \& Green 1994).  In addition, Richards et~al. (2002)
suggested that there may be a relationship between the line shape,
equivalent width, and blueshift of the high-ionization (e.g., $\CIV$)
lines with respect to the low-ionization (e.g., $\MgII$) lines of the
quasar.  Vanden Berk et~al. (2003) confirmed the findings of Richards
et~al. (2002), and also noted that the Baldwin Effect appears to
evolve with redshift.

Our fitting process is designed to take advantage of the general
similarity of quasar UV/optical spectra with a general first-pass fit,
while accounting for the differences arising from line shape,
luminosity, high-ionization line blueshift, and redshift.  We begin by
fitting to a geometric mean composite spectrum constructed using the
methods of Vanden Berk et~al. (2001) from the quasars in the SDSS
First Data Release (DR1; Abazajian et~al. 2003) quasar catalog
(Schneider et~al. 2003).  This general first-pass fit is designed to
provide initial estimates of spectral index and reddening and to weight
out absorption regions from the final fitting process (see \S 3.2
below).  We account for differing luminosities, line widths (and thus
indirectly any mean blueshift of $\CIV$ with respect to $\MgII$) and
possible redshift trends by finding the best-fit template for each
quasar from a set of templates binned in line width, luminosity, and
redshift.

Quasars for the set of template spectra were selected from the
SpecPhotoAll table of the public DR3 Catalog Archive Server (Stoughton
et~al. 2002b), requiring a redshift confidence of 95\% or greater and
rejecting inconsistent or manually corrected
redshifts\footnote{Inconsistent or manually corrected redshifts are
rejected by requiring {\tt zStatus in \{3,4,6,7,9\}} in the public
SDSS DR3 Catalog Archive Server.}.  We also required either the $\CIV$
or $\MgII$ line to have been successfully measured by the automated
Gaussian-fitting pipeline, with $\chi^2_{\nu}<2$, ${\rm EW} /
\sigma_{\rm EW} > 10$ and ${\rm FWHM} / \sigma_{\rm FWHM} > 5$.
Objects with a null or negative $i$-band magnitude measurement which
prevented calculation of the $i$-band absolute magnitude were
excluded.  Finally, objects with known BALs and strong narrow
absorption systems (as identified by various SDSS workers during
visual inspections) were excluded to ensure that each template
reflected an unabsorbed quasar spectrum.
%All spectra were corrected for Galactic reddening according to the
%maps of Schlegel, Finkbeiner, \& Davis (1998) before constructing the
%final set of templates.

\begin{figure}
\plotone{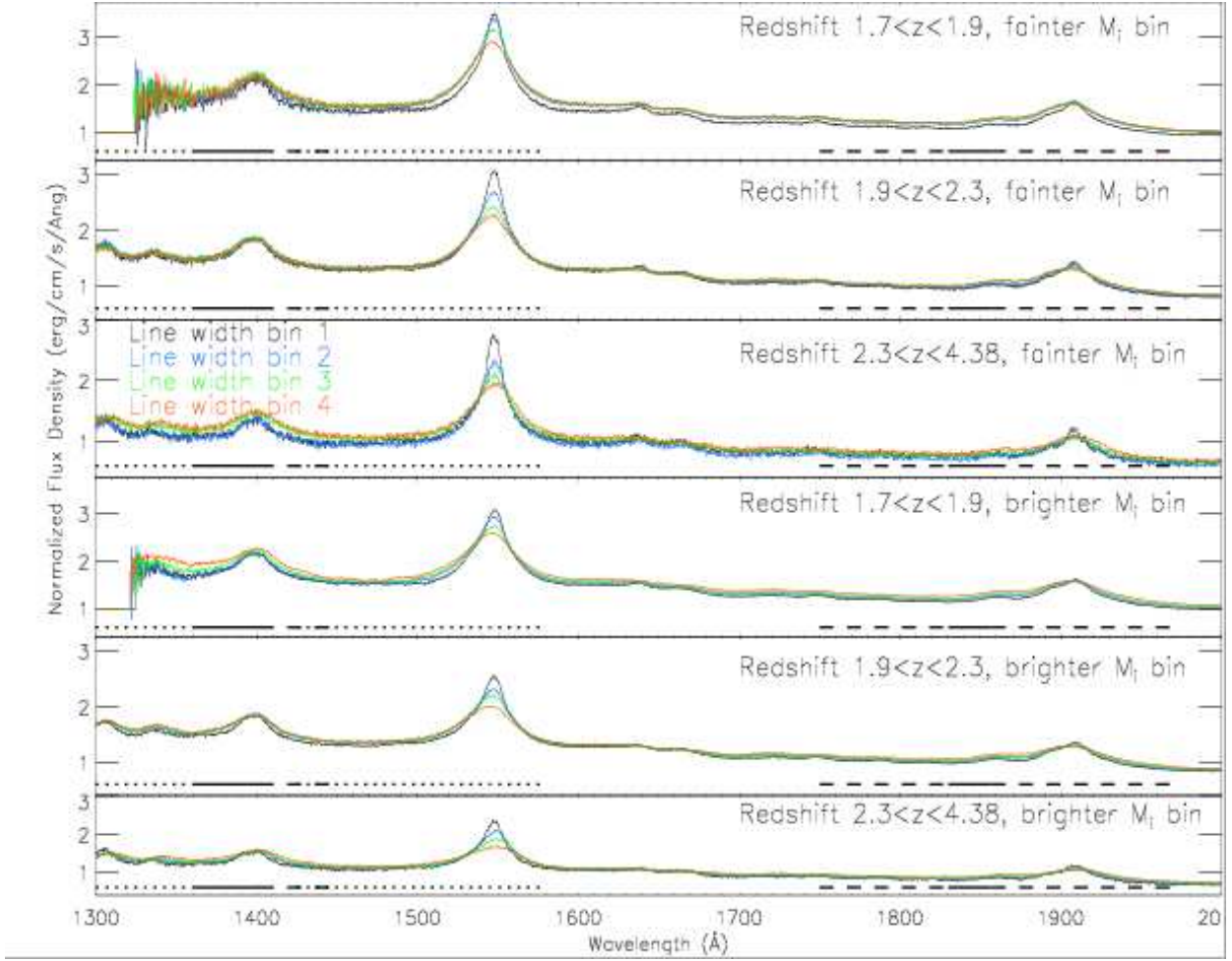}
\caption{The templates used in fitting for $\CIV$ absorption, as
  described in Table \ref{tbl:templates}.  The four line width bins
  are shown by different colors in each plot, from narrowest to
  broadest.  Each set of three redshift bins goes down the page with
  increasing redshift.  The top three plots are in the less luminous
  absolute i-band bin.  The templates are plotted for the $\CIV$
  absorption region and include the $1710 \pm 10$ \AA\ normalization
  window.  Below each spectrum are lines representing the weights: the
  solid lines represent $w(\lambda)=0$ regions, the dashed lines are
  $w(\lambda)=0.5$, and the dotted lines are the special $\CIV$ and
  $\SiIV$ absorption regions, according to Table \ref{tbl:weights} and
  \S 3.2.  The spectral resolution is $\approx
  2000$. \label{fig:civtemp}}
\end{figure}

\begin{figure}
\plotone{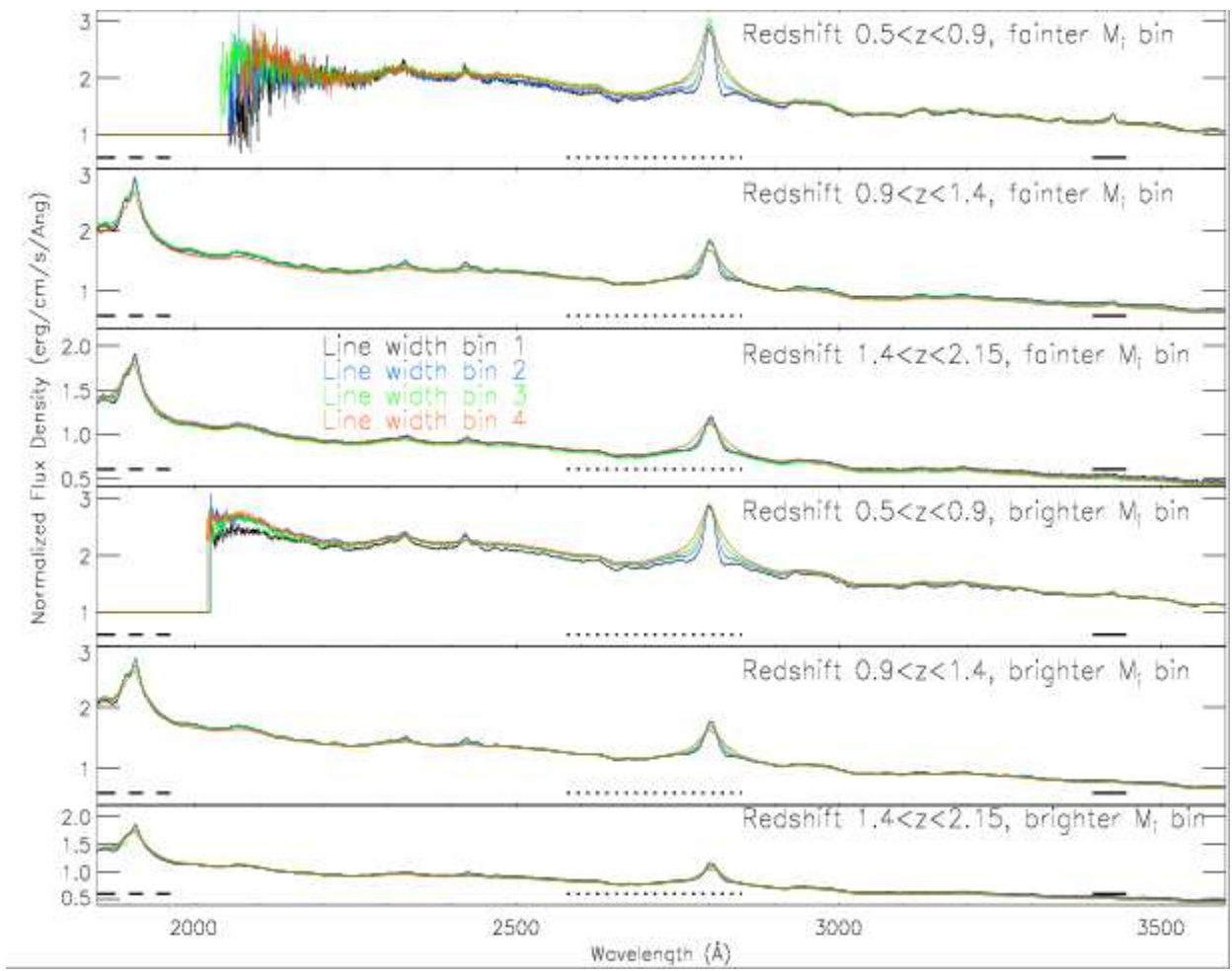}
\caption{The templates for $\MgII$ absorption, as described in Table
  \ref{tbl:templates}.  As in Figure \ref{fig:civtemp}, each plot has
  four line-width bins and the top three plots, descending with
  increasing redshift, are in the less luminous absolute i-band bin.
  The templates extend over the $\MgII$ absorption region and include
  the $2910 \pm 10$ \AA\ normalization window.  Below each spectrum
  the solid lines represent $w(\lambda)=0$ regions, the dashed lines
  are $w(\lambda)=0.5$, and the dotted lines are the special $\MgII$
  emission and absorption region, according to Table \ref{tbl:weights}
  and \S 3.2. \label{fig:mgiitemp}}
\end{figure}

The properties of the final-fit template samples are shown in Table
\ref{tbl:templates}.  Composites for $\CIV$ were constructed in three
redshift ranges: $1.52<z<1.9$, $1.9<z<2.3$ and $2.3<z<4.90$.  Within
each redshift range, quasars were separated into the upper and lower
halves of the absolute $i$-band luminosity distribution.  These
absolute magnitude bins were further subdivided into quartiles of
$\CIV$ line width (narrowest, narrow, broad, and broadest).  Thus,
eight templates were created for each of the three redshift bins.  The
templates are enumerated in our discussion and final catalog as
templates 0-3 representing the less luminous $M_i$ bin in order of
narrowest line width to broadest, and templates 4-7 representing the
more luminous $M_i$ bin in the same order of line width.  An identical
approach was used to create $\MgII$ templates in the three redshift
ranges $0.39<z<0.9$, $0.9<z<1.4$ and $1.4<z<2.2$.  Quasars with
redshift $1.7 \le z \le 2.15$ have two best-fitting templates, one for
$\CIV$ and one for $\MgII$.  The template spectra for $\CIV$ are shown
in Figure \ref{fig:civtemp} and those for $\MgII$ in Figure
\ref{fig:mgiitemp}.

\subsection{Fitting the Continuum}

The ultraviolet/optical continua of quasars are generally
well-described by a power-law $f_{\nu} \propto \nu^{\alpha_{\nu}}$,
where $\alpha_{\nu}$ is the spectral index (e.g, Richstone \& Schmidt
1980; Vanden Berk et~al. 2001).  However, a significant fraction of
quasars show evidence of dust reddening (e.g, Richards et~al. 2003,
Hopkins et~al. 2004), which must be included in any automated
continuum fitting procedure.  For simplicity we assume that all dust
reddening beyond the Milky Way value occurs at the quasar redshift.
Our automated fitting algorithm works in wavelength space and allows
both the spectral index and reddening to vary, using Newton-Raphson
minimization to fit the template to each sample spectrum.

The reddening can be described by the dust extinction $A_{\lambda}$,
\begin{equation}
A_{\lambda} = E (\lambda-V) + R_V \times E (B-V).
\end{equation}
Here $E(B-V)$ is the color excess and we adopt $R_V=2.93$ (as in Pei
1992).  In general, dust extinction laws are roughly proportional to
$\lambda^{-1}$, but there are differences in the exact slope of the
extinction and in the strength of the ``2175 \AA\ bump'' (Sprayberry
\& Foltz 1992).  Extragalactic sources have a weak or missing 2175
\AA\ bump, so we use the Small Magellanic Cloud reddening law of
Prevot et~al. (1984) as implemented by Pei (1992).
%which has the smallest 2175 \AA\ bump of the reddening curves.  
%We represent the SMC reddening law applied to the quasar spectrum as
%\begin{equation}
%f_{\lambda}(E) \propto 10^{-aE\xi(\lambda)}
%\end{equation}
%with $a=0.4(1+R_V)$, $R_V=2.93$, and $\xi(\lambda)$ the extinction
%curve given by Pei (1992).

Each template spectrum $f_{t,\lambda}$ (with intrinsic spectral index
$\alpha_{\lambda t}$ and reddening $E_t$) is fit by adjusting the
spectral index $\alpha_\lambda$ and reddening $E \equiv E(B-V)$ to
obtain the fitted template spectrum $f_{t',\lambda}$:
\begin{equation}
f_{t',\lambda}(\alpha_\lambda,E) = f_{t,\lambda} \lambda^{\alpha_\lambda-\alpha_{\lambda t}} 10^{-a(E-E_t)\xi(\lambda)}.
\end{equation}
where $a=0.4(1+R_V)$, $R_V=2.93$, $\xi(\lambda)$ is the extinction
curve, and $\alpha_\lambda \equiv -2-\alpha_{\nu}$ is the spectral
index.  The best-fit template is found by minimizing the chi-squared
function
\begin{equation}
\chi^2(\alpha_\lambda,E) = \frac {\sum_{\lambda} \left[ \frac{f_{t',\lambda}(\alpha_\lambda,E)-f_i(\lambda)}{\sigma_i(\lambda)} \right]^2 w(\lambda)} {\sum_{\lambda} w(\lambda)}
\end{equation}
where the input spectrum is $f_i(\lambda)$, its error is
$\sigma_i(\lambda)$, and $w(\lambda)$ is our weighting function.

Our weighting function $w(\lambda)$ is designed to limit the influence
of interpolated pixels, emission line regions, and absorption regions.
Interpolated pixels are identified by the error of zero assigned by
the pipeline and are given a weight $w(\lambda)=0$.  Weighting the
emission line regions and absorption regions is more complicated; see
Table \ref{tbl:weights}.  Unweighted line regions include the $\Lya$
forest, wavelengths redward of 4050 \AA\ (a single power-law continuum
is not a good fit past 4100 \AA; see Figure 6 of Vanden Berk
et~al. 2001), the narrow emission lines $\NeV$, $\OII$, and $\NeIII$,
and the blue halves of the $\NV$, $\SiIV$/$\OIV$ and $\CIII$ emission
lines.  Because the widths of the emission lines are important in
template selection, $w(\lambda)=0.5$ is assigned to the red halves of
strong emission lines, where intrinsic absorption should not
significantly affect the emission profile.  Accurately fitting the
$\CIV$ and $\MgII$ emission regions is especially important, so we set
$w(\lambda)=0.5$ within $\pm 5000~\kms$ of the emission line and
weight out absorption separately (see below).

Absorption regions in a sample spectrum cause the flux density in the
fitted template to be under-estimated.  To eliminate these regions
from the fitting, we identify absorption from the first-pass fit and
then weight out those regions in the final, multiple-template fitting.
We identify absorption only from $-5000~\kms$ to $29000~\kms$ relative
to the rest wavelengths of the $\CIV$ and $\MgII$ emission
lines.\footnote{For some quasars, we adjusted the redshift used for
the template fitting from the value given in the DR3 quasar catalog to
better reflect the redshift of the $\CIV$ or $\MgII$ line.  This is
necessary because individual emission lines in quasars can have large
velocity shifts relative to the mean emission-line redshift.}  Since
poor fits to an emission line can masquerade as absorption troughs,
and since our first-pass fit to the spectrum does not fit the emission
line shapes, within $\pm 5000~\kms$ of the $\CIV$ and $\MgII$ rest
wavelengths we identify absorption relative to a simple linear
continuum.  Flux density values $\ge 2\sigma$ below this linear
continuum, or $\ge 4\sigma$ below the continuum in the region
$5000~\kms$ to $29000~\kms$ shortward of $\CIV$ or $\MgII$, are
assigned $w(\lambda)=0$.  Because the $\CIV$ and $\SiIV$ absorption
regions often exhibit similar absorption features, we apply the $\CIV$
absorption weighting to the same velocities in the $\SiIV$ region
($-5000~\kms$ to $29000~\kms$ from 1402~\AA).  The first-pass fit also
provides an initial guess on the values of $\alpha_\lambda$ and $E$
for the multiple-template fitting.

As the values of $\alpha_\lambda$ and $E$ are varied, the template
spectrum is scaled to the quasar spectrum before the $\chi^2$ function
is computed.  Scaling is performed by calculating the average flux
density of each spectrum over small normalization windows.  For $\CIV$
this window is $1710 \pm 10$ \AA, a reasonable continuum region
(Vanden Berk et al. 2001).  For $\MgII$ BAL identification we use the
window $2910 \pm 10$ \AA, the first minimum in the $\FeII$ flux
redward of $\MgII$ (Vestergaard \& Wilkes 2001).  A normalization
window around 3020 \AA\ might be better (Sigut, Pradhan, \& Nahar
2004), but would limit the LoBAL sample to lower redshifts.  We choose
this region over the two separate normalization regions of Reichard
et~al. (2003a) for self-consistency.
%If extreme noise or interpolation in the
%normalization region causes the average flux density to be negative,
%the template spectrum is scaled to zero and the object is inspected
%later for manual template fitting.
%PBH: mentioned later, in manual fitting section.

The $\chi^2$ function in Equation 3 is minimized using the
quadratically converging modified Newton-Raphson method used by
Reichard et al. (2003a).  The initial values of the spectral index and
reddening are from the DR1 quasar composite spectrum: $E=0$ and
$\alpha_\lambda=-1.55$ (corresponding to $\alpha_\nu=-0.45$).  The
first-pass quasar template spectrum described above is then fit to
each quasar spectrum, and the best values of $\alpha_\lambda$ and
$E$ from this fit are used as the initial spectral index and reddening
for each template fit.  If the Newton-Raphson method does not converge
in 50 iterations, the $\chi^2$ function is evaluated over a wide grid
of $(\alpha_\lambda,E)$ to find the minimizing values.  This
occurred for only 4 $\CIV$ and 49 $\MgII$ objects.  The grid method is
computationally intensive and was used only as a back-up approach.

Because the absorption index starts at the zero velocity shift from
the emission line peak, the emission lines of the quasar spectra need
to be fit well for an accurate measure of the absorption index.  We
employ independent emission line scaling for $\CIV$ and $\MgII$, which
is justified because we are not seeking a physical characterization of
the quasar spectrum in the fitting process, merely a representative
continuum with which to identify absorption troughs.  After the
power-law and reddening fit and the continuum scaling, a linear
continuum is subtracted from both emission regions (1494-1620 \AA\ for
$\CIV$ and 2686-2913 \AA\ for $\MgII$).  For both lines the template
emission line profile is scaled to match the peak flux in the
5-pixel-smoothed sample emission region and the continuum is then
re-added to the scaled emission line profile.

Despite our best efforts at automating our template-fitting methods,
some objects are poorly fit by all of the templates.  A bad template
fit often results in a falsely identified or missed BALQSO.  We
inspect all quasars in our sample and manually tweak our templates to
fit in the $\CIV$ or $\MgII$ regions where the fitting fails
spectacularly and results in a mis-identification.  We had to manually
fit the template for 347 quasars in the $\CIV$ region and 3 in the
$\MgII$ region.  (There are 20 times more $\CIV$ BALs than $\MgII$
BALs in the catalog, so we would have expected 17 $\MgII$ BALs needing
adjustment.  The difference may be because the spectral region about
$\MgII$ is easier to fit than the region about $\CIV$.)
%Most often, the manual adjustment to
%the template requires only re-scaling the template to account for
%unusual emission line profiles or noisy normalization regions in the
%sample.	%PBH: redundant w/later discussion

\section{Selection Criteria of BALQSOs and Traditional BALQSOs}

The biggest challenges in identifying intrinsic absorption in quasars
are avoiding intervening and host systems and accounting for noise.
Completely eliminating host and intervening absorption is impossible
without high-resolution spectra, but most systems of this type can be
eliminated by defining a minimum width for absorption to be considered
an intrinsic trough.  To avoid contamination by noise, we must define
a minimum depth.  We now examine the historical treatment of this
issue, and suggest a modified classification system appropriate for
the SDSS.

\subsection{The Traditional Balnicity Index}

Weymann et~al. (1991) defined BALs to be at least $2000~\kms$ wide and
at least 10\% below the continuum at maximum depth.  These objects
were quantified by the ``balnicity index'' (BI), a sum of the
``modified equivalent width'' of the portions of all contiguous BAL
troughs between $3000~\kms$ and $25000~\kms$ shortward of 1549 \AA.
The equivalent width is ``modified'' because only those parts of the
troughs beyond the first $2000~\kms$ of width that dip below 10\% of
the continuum are included.  This is defined:
\begin{equation}
{\rm BI} = \int_{3000}^{25000} \left[ 1 - \frac{f(v)}{0.9} \right] C~dv
\end{equation}
where $f(v)$ is the normalized flux, $C=1$ at trough velocities more
than $2000~\kms$ from the start of a contiguous trough, and $C=0$
elsewhere.  The BI can take on any value in the range $0 \le {\rm BI}
\le 20000~\kms$.  The formal error on the balnicity index is
\begin{equation}
\sigma_{\rm BI}^2 = \int_{3000}^{25000} \left( \frac{\sigma_{f(v)}}{0.9} \right)^2 C~dv
\end{equation}
but uncertainties in the continuum usually dominate over this term.

Weymann et~al. (1991) selected their minimum width and depth to avoid
host and intervening systems at their spectral resolution, as well as
noise in the data.  With our multiple-template fitting technique and
somewhat higher resolution spectra, we may be able to reduce the
minimum width and/or depth in our definition of intrinsic absorption.
While we retain the Weymann et~al. (1991) definition of traditional
BALQSOs as quasars with ${\rm BI}>0$, we also present a modification
of the BI which includes all absorption.

\subsection{Testing Definitions of Broad Intrinsic Absorption}

The ideal definition of broad intrinsic absorption was discussed in
the appendix to Hall et~al. (2002).  It would include shallow troughs,
``mini-BALs'' which are not as wide as BALs yet may share all the rest
of their properties, and absorption within $3000~\kms$ and beyond
$25000~\kms$ of the systematic velocity as long as it was not confused
with troughs from other transitions.
%(e.g., $\CIV$ absorption would be identified blueshifted as far as the
%$\SiIV$ emission and absorption region).
%
This ideal definition of broad intrinsic absorption requires perfect
knowledge of the systematic velocity, the continuum, and which troughs
are truly intrinsic and which arise from host or intervening systems.
The automated SDSS redshift determination scheme should produce
consistent systemic velocities, and
%the use of templates binned in line width space allows
%acceptable fitting of any blueshifted lines in the spectrum.  Our
our system of continuum fitting with multiple templates is designed to
address the problem of accurately estimating the continuum.  However,
we find two major difficulties in continuum fitting which require
special attention: emission line regions and interpolated regions.

Both the $\CIV$ doublet (separation $497~\kms$) and the $\MgII$
doublet (separation $768~\kms$) almost always appear as a single broad
emission line in quasar spectra.  A minimum trough width of $450~\kms$
would ideally exclude all $\CIV$ features wherein the doublet was
still distinct.  Velocity spreads over $\sim 450~\kms$ are very rare
for intervening $\CIV$ systems and coincidences of multiple systems on
these scales are uncommon because intervening narrow $\CIV$ absorbers
are uncorrelated on velocity separations $>400~\kms$ (Churchill et
al. 2000, Ding et~al. 2003), so this minimum width would also exclude
most intervening systems.  Following the same reasoning, we arrive at
a minimum intrinsic absorption width of $750~\kms$ for $\MgII$.

However, the noise and variable resolution of our spectra, as well as
poor continuum fits, might introduce false identifications if we adopt
too narrow a minimum width.  For example, if the $\CIV$ doublet was
blended, we would not expect absorption narrower than $1000~\kms$;
similarly, we would not expect absorption narrower than $1500~\kms$
for blended $\MgII$ doublets.  We tested the lower three of these
possible minimum widths: $450~\kms$, $750~\kms$, and $1000~\kms$.

In addition to the minimum width of the trough, we must also define a
minimum depth.  If we define a shallow minimum depth, we risk
identifying false troughs from noise and poor continuum fits, but if
we choose a minimum depth too deep, we risk missing weak troughs.
Weymann et~al. (1991) and Reichard et~al. (2003a) used a minimum depth
of 10\% below the continuum.  A $3\sigma$ minimum depth definition
might be more reasonable for a sample which includes spectra with very
different signal-to-noise ratios, but may result in different answers
for repeat observations.  We tested both 10\% and $3\sigma$ minimum
depths using the absorption index (AI) defined in \S 4.3.

\begin{figure}
\plotone{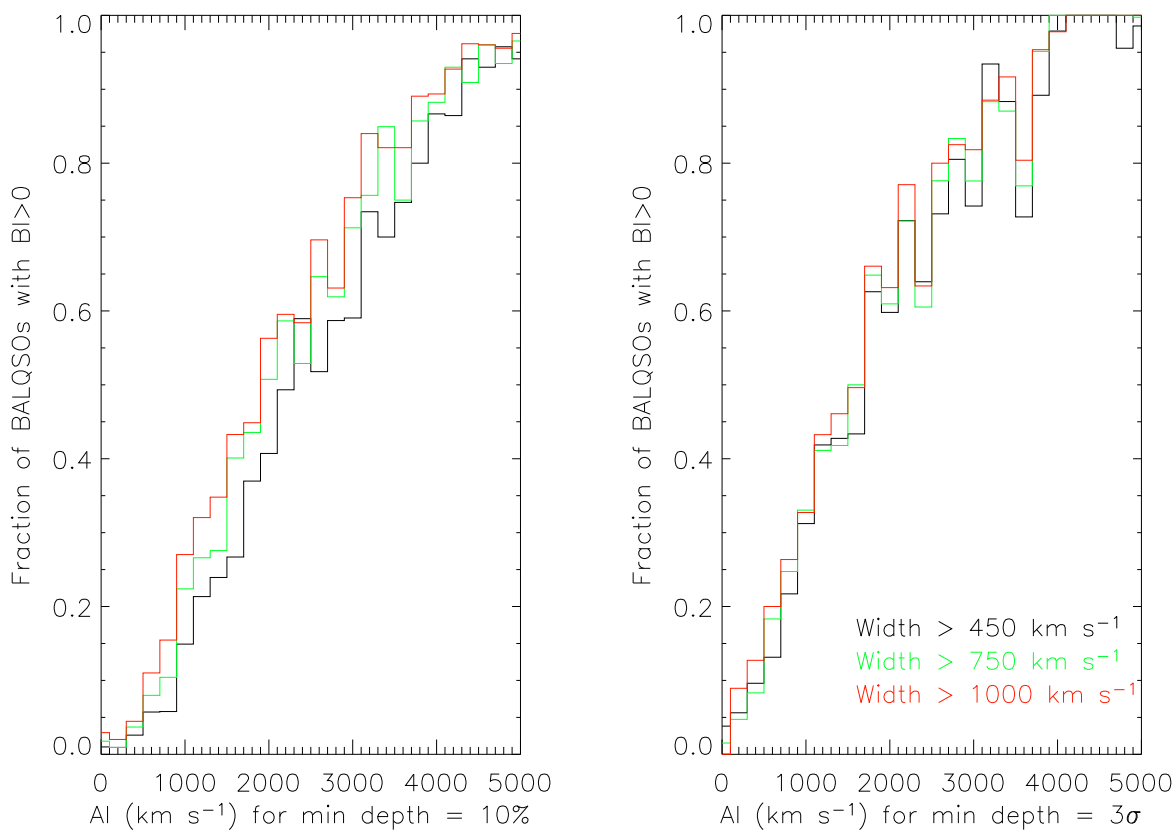}
\caption{The AI distribution of BALQSOs which also have ${\rm BI}>0$
  for varying minimum widths.  On the left is shown the distribution
  for a minimum depth of 10\% below the continuum, and on the right
  for a minimum depth of 3$\sigma$.  Each bin of AI is $200~\kms$
  wide.  Note that the last bin includes all quasars of AI $\ge
  5000~\kms$.
\label{fig:widedeep}}
\end{figure}

Among the 16883 SDSS DR3 quasars in the $\CIV$ redshift range $1.7 \le
z \le 4.38$, using the traditional 10\% minimum depth, 60.6\% have a
nonzero absorption index (as defined in \S 4.3) for a minimum width of
$450~\kms$, 44.7\% for a minimum width of $750~\kms$, and 32.3\% for a
minimum width of $1000~\kms$.  With a $3\sigma$ minimum depth, the
BALQSO fractions decrease significantly: 31.0\% for a minimum width of
$450~\kms$, 26.4\% for $750~\kms$, and 20.2\% for $1000~\kms$.  In
comparison, only 12.0\% of the 16883 quasars are identified as
traditional BALQSOs.  The number of quasars with nonzero BI for each
binned value of AI are shown in Figure \ref{fig:widedeep} for the
different minimum widths and depths.  Very few of the BALQSOs
identified with a minimum depth of $450~\kms$ also have ${\rm BI}>0$.
Because so many fewer quasars are identified for an error-dependent
minimum depth of $3\sigma$, we choose to retain the 10\% minimum depth
of the traditional balnicity.

\begin{figure}
\plotone{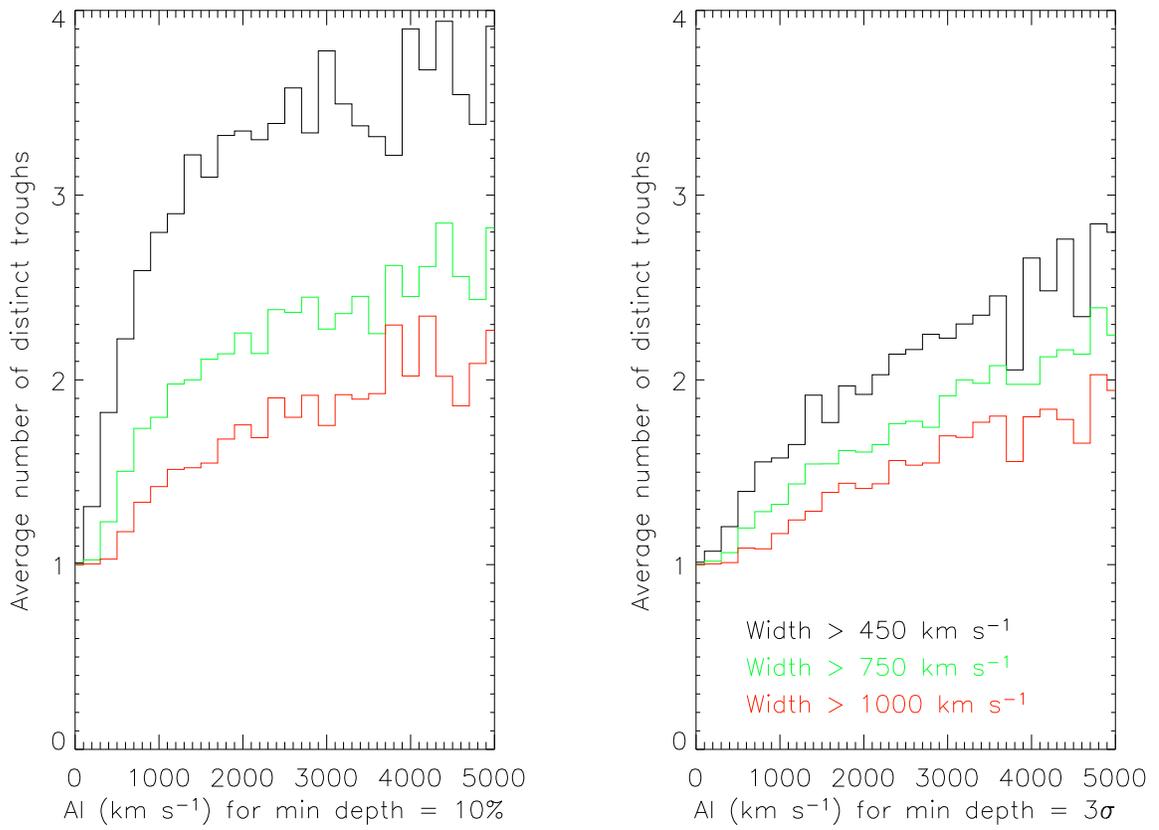}
\caption{The average number of troughs in each BALQSO for given values
  of AI, calculated with different minimum depths and widths.  A large
  number of troughs on average may indicate the identification of too
  many narrow troughs due to host absorption, intervening systems,
  and/or noise.  Each bin of AI is $200~\kms$
  wide. \label{fig:troughs}}
\end{figure}

A large number of distinct troughs could be attributed to intervening
absorption systems and noise, and thus counting the number of troughs
for each minimum width might provide a measure of the appropriateness
of each minimum width.  Figure \ref{fig:troughs} shows the number of
troughs identified for each minimum width for both minimum depths.  In
both figures we see that a large number of troughs are identified
using minimum widths of $450~\kms$ and $750~\kms$.  We conclude that
the minimum widths of $450~\kms$ and $750~\kms$ are too aggressive for
our quasar sample, and choose a minimum width of $1000~\kms$ for our
definition of BALQSOs.

\subsection{The Absorption Index}

The ``absorption index'' (AI) was first proposed by Hall et
al. (2002).  In this paper, we modify that definition slightly so that
AI is a true equivalent width, measuring all absorption within the
limits of every trough, and so that the integration limit extends to
$29000~\kms$.  Thus our AI is defined as:
\begin{equation}
{\rm AI} = \int_{0}^{29000} \left[ 1 - f(v) \right] C'~dv
\end{equation}
where $f(v)$ is the normalized template-subtracted flux.  The variable
$C'(v)$ has the value $C'=0$ except in contiguous troughs which exceed
the minimum depth (10\%) and the minimum width (1000\,\kms), in which
case $C'=1$.
%\begin{enumerate}
  %\item the trough exceeds the minimum width (in the end we adopt 1000~\kms)
  %\item the trough exceeds the minimum depth over the entire trough (in the end we adopt 10\%)
%%\item %$\chi_{\rm trough}^2 = \sum {{1}\over{N}} \left( \frac{1-f(v)}{\sigma}
    %%%\right)^2 > 10$: 
%%the reduced chi-squared of the trough is greater than 10
  %%\item if $v < 5000~\kms$, $f_{\rm flat}(v)-f(v) \ge 2\sigma$ for
    %%\emph{any} $v$ in the range $\Delta v$: the trough must drop at
    %%least $2\sigma$ below the linear continuum $f_{\rm flat}(v)$ at
    %%least once, as described above.
%\end{enumerate}
Zero velocity is defined using the SDSS redshift and the longest
wavelength line of the doublet: 1550.77 \AA\ for $\CIV$ and 2803.53
\AA\ for $\MgII$.\footnote{Because the SDSS redshift is based on
emission lines which can be blueshifted from the true systemic
velocity, absorption can sometimes appear redward of zero velocity as
defined above.  We do not catalog such absorption unless it extends
1000\,\kms\ shortward of the zero velocity wavelength.}  The minimum
AI is $100~\kms$, and the maximum is $29000~\kms$.  The formal error
of the absorption index is given by:
\begin{equation}
\sigma_{\rm AI}^2 = \int_{0}^{29000} \sigma_{f(v)}^2 C'~dv
\end{equation}
Again, the error introduced by the continuum placement will
dominate over this error term in most cases.

Even the most appropriate minimum depth and width may not successfully
remove all contamination by noise-induced false troughs.  For this
reason we also calculated the reduced chi-squared of each trough:
\begin{equation}
\chi_{\rm trough}^2 = \sum {{1}\over{N}} \left( \frac{1-f(v)}{\sigma} \right)^2
\end{equation}
Here $N$ is the number of pixels in the trough and $f(v)$ is the
normalized template-subtracted flux ($f(v)=1$ where the sample flux
and the template are equal, and $f=0$ for a black trough).  The
greater the value of $\chi_{\rm trough}^2$, the more likely the trough
is not due to noise.  We require that troughs have $\chi_{\rm
trough}^2 > 10$ to be considered true BALQSO troughs.  After visual
inspection, we find that 34 quasars with very low signal-to-noise
ratio spectra have troughs which are wrongly eliminated by the
$\chi_{\rm trough}^2 > 10$ cut.  We make an exception and do not
require $\chi_{\rm trough}^2 > 10$ for these 34 objects.

During the initial stages of BALQSO identification, we encountered a
large number of false identifications within $5000~\kms$ of the
emission line peak.  This is caused by the difficulty of accurately
fitting, in every quasar, the exact profile of the $\CIV$ or $\MgII$
emission line.  To avoid these false identifications we employ an
extra minimum depth requirement in the emission line region.  First we
define a linear continuum across the emission region of the template
which is an interpolated line across $\pm5000~\kms$ of the emission
center.  We then require absorption in this region to drop below the
linear continuum by at least twice the noise level at least once
within a trough, in addition to fulfilling the $1000~\kms$ minimum
width and $10\%$ minimum depth requirements.  This method successfully
repressed most of the false identifications in the line region.

Interpolated pixels are weighted out of the continuum-fitting process,
but can still result in the misidentification of BAL troughs.
Interpolation most often occurs in noisy areas of the spectrum, so it
is possible that if the noise causes the two endpoints of an
interpolated region wider than $1000~\kms$ to lie below the minimum
depth, a BAL trough will be identified.  To avoid these false
identifications, we ignore all interpolated pixels during BAL trough
identification.  Recall from \S 2 that we extend the SDSS
pipeline-interpolated regions by three pixels on either side.

\begin{figure}
\plotone{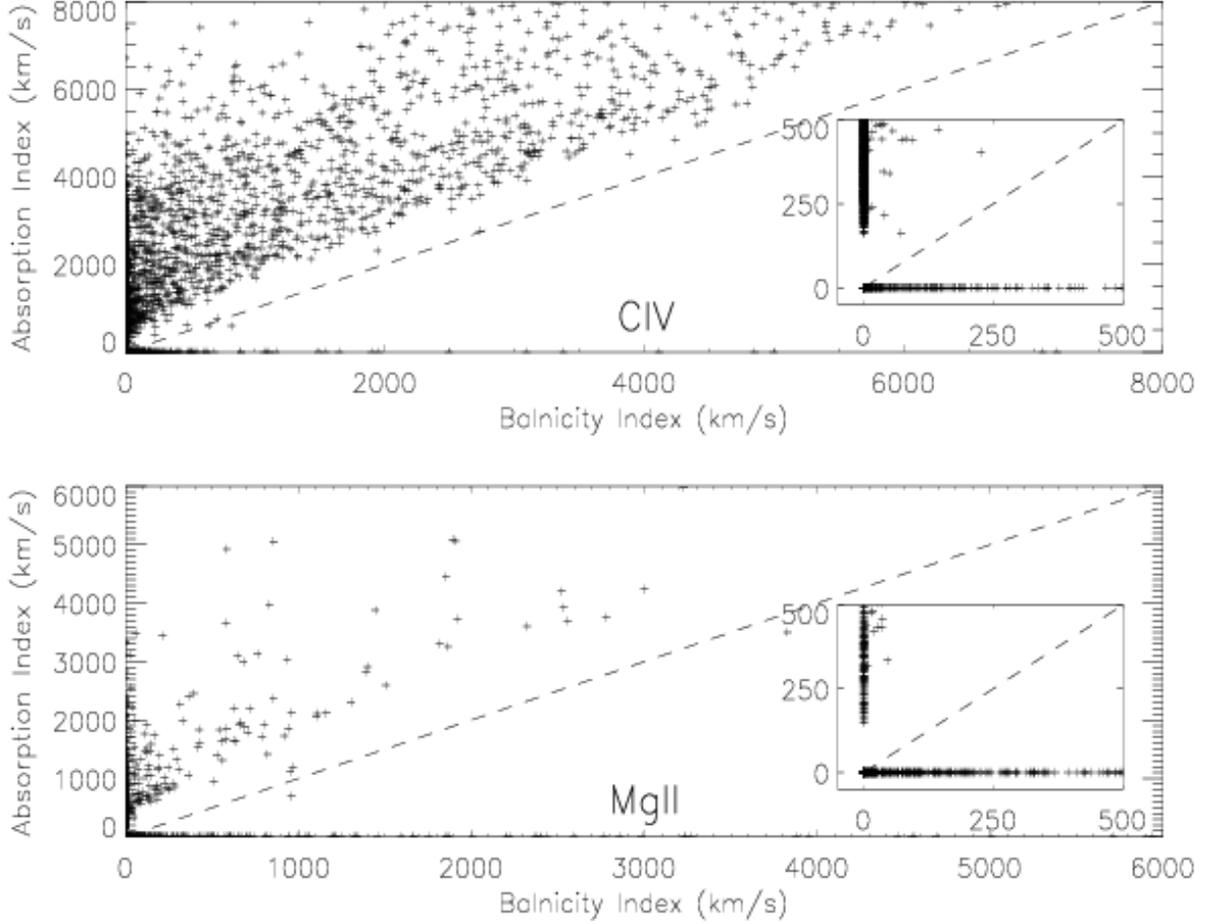}
\caption{The absorption index vs. the balnicity index for the 5418
  $\CIV$ BALQSOs (top) and 3032 $\MgII$ BALQSOs (bottom) in our
  catalog.  Except for a handful of objects, the AI is greater than
  the BI because it includes all absorption, within the minimum depth
  and width.  The AI is also significantly greater than the BI for
  many objects because it identifies narrower troughs and troughs near
  the zero velocity blueshift.  The regions of low AI and BI are shown
  as insets.  A few objects are identified to have ${\rm AI}<{\rm BI}$
  because of our linear continuum restriction in the emission line
  region and our $\chi_{0}^2 \ge 10$ requirement for each trough in
  the AI calculation (see \S 4.3). \label{fig:AIvsBI}}
\end{figure}

We use the AI to identify both HiBALs with $\CIV$ absorption and
LoBALs with absorption shortward of the $\MgII$ through our automated
techniques.  Figure \ref{fig:AIvsBI} shows the general differences
between quantifying our objects with the AI and BI.  In most objects,
${\rm AI}>{\rm BI}$ because the AI includes all absorption in its
definition and often identifies troughs which are too narrow to be
identified using the BI.  However, if a trough identified by the BI
calculation has $\chi_0^2<10$ or is not deep enough within $5000~\kms$
of the line region, it is not considered a trough in the AI
calculation.  The automated fitting identified 138 objects with ${\rm
BI}>{\rm AI}$ in the $\CIV$ region, and 137 ${\rm BI}>{\rm AI}$
objects in the $\MgII$ region.  Visual inspection confirmed that these
objects have only spurious troughs; therefore, they are not part of
our BAL catalog.  For the differences between the AI and BI for
individual objects, see Figures \ref{fig:BALQSO} and \ref{fig:LoBAL},
as well as the accompanying discussion in \S 5.1 and 5.2.

\subsection{BALQSO Selection Summary}

A summary of our selection of BALQSOs is presented in the list below.

\begin{enumerate}
  \item From the DR3 QSO catalog, select all objects of 
    $0.5\le z\le 2.15$ for studies of absorption in the $\MgII$ region,
    and quasars of $1.7\le z\le 4.38$ for the $\CIV$ region.
  \item Weight emission lines and interpolated pixels according to
    Table \ref{tbl:weights}.
  \item Fit the DR1 quasar composite to the quasar, using the spectral
    index, reddening, and scale as free parameters.  Weight out
    regions of absorption, identified as flux $\ge 4\sigma$ below the
    template in the region $\pm5000~\kms$ about the line center, and
    $\ge 2\sigma$ below the template in the region $5000~\kms$ to
    $29000~\kms$.
  \item Find the best-fit template from a set of 8 templates in the
    redshift bin.  The templates are described in Table
    \ref{tbl:templates} and shown in Figures \ref{fig:civtemp} and
    \ref{fig:mgiitemp}.  The spectral index, reddening, and scale are
    free parameters in the fitting.
  \item Independently scale the $\CIV$ or $\MgII$ emission line
    regions.
  \item In the spectral region $0~\kms$ to $29000~\kms$ from the
    emission line center, identify regions of flux at least 10\% below
    the continuum and $1000~\kms$ wide.  Ignore interpolated pixels.
  \item Eliminate from consideration any troughs less than $5000~\kms$
    from the emission line center that do not have at least one pixel
    $>2\sigma$ below a linear continuum across the emission region.
  \item Eliminate from consideration any troughs with $\chi_0^2<10$
    between the template and the quasar.  (We relax this criterion for
    25 objects with low signal-to-noise ratio spectra.)
  \item Calculate the AI and BI.
  \item Inspect all objects.  Manually adjust the best-fit templates
    for quasars with poor automatic fits and repeat steps 6-9.

\end{enumerate}

\subsection{Other metrics}

In addition to the measures of equivalent width provided by the
absorption index and the traditional balnicity, other BALQSO
quantities can be measured (e.g., Lee \& Turnshek 1995).  We
calculated a number of additional metrics to learn which have the most
relevance in the characterization of spectral absorption troughs:

\begin{itemize}
    \item number of troughs
    \item $v_{\rm min}$ and $v_{\rm max}$ for each trough
    \item average fractional depth for each trough, given by
      $\Gamma_{\rm avg} = \frac{\rm AI_{\rm trough}}{(v_{\rm max} -
      v_{\rm min})}$
    \item maximum fractional depth and maximum width of each trough
      and over the entire absorption region
    \item reduced chi-squared of the trough, given by $\chi^2 = \sum
      {{1}\over{N}} \left( \frac{1-f}{\sigma} \right)^2$, with $N$ the
      number of pixels in the trough
    \item weighted average velocity for each trough and over the
      entire absorption region, given by $v_{\rm wtavg} =
      \frac{\int_{0}^{29000} v [1-f(v)] C' dv} {\rm AI}$
    \item second moment for each trough, given by $\sigma_{\rm rms}^2
      = \frac{\int_{0}^{29000} (v-v_{\rm wtavg})^2 [1-f(v)] C' dv}
      {\rm AI}$
\end{itemize} These metrics were measured on all objects of ${\rm
AI}>0$.  Discussion of metrics relevant to individual troughs are not
presented here but will be discussed in a future paper (Hall
et~al. 2006).

\section{BALQSO Catalog}

%We categorize quasars with broad absorption lines into four
%overlapping categories: (1) traditional HiBALs, as defined by Weymann
%et~al. (1991) with ${\rm BI}>0$ in the $\CIV$ absorption region, (2)
%HiBALs, with ${\rm AI}>0$ in the $\CIV$ region, (3) LoBALs, with ${\rm
%AI}>0$ in the $\MgII$ region, and (4) FeLoBALs, identified only
%through visual inspection.  Examples of each type of object are
%discussed below.  The frequencies of BAL phenomena given in the
%following subsections are raw only; a future paper will address 
%the true frequencies, after correction for selection effects. 
%
%Our entire catalog of HiBALs, LoBALs, and FeLoBALs is available as an

Our entire BALQSO catalog is available electronically.  Each column of
the electronic table is described in Table \ref{tbl:catdesc}.  We
present the first page of the catalog in tabular format in Table
\ref{tbl:catalog}.  We also include a supplementary Table of purely
visually identified $z>4.38$ BALQSOs in the Appendix.

The BAL subtype is coded in the catalog as follows: Lo denotes a LoBAL
with ${\rm AI}>0$ in the $\MgII$ region; LoF denotes a FeLoBAL with
${\rm AI}>0$ in the $\MgII$ region; HLF denotes a FeLoBAL with ${\rm
AI}>0$ in the $\CIV$ region; HL denotes a HiBAL in which broad
($\geq$1000~\kms) low-ionization absorption of \ion{Mg}{2},
\ion{Al}{3} or \ion{C}{2} is also seen by visual inspection; Hi
denotes a HiBAL-only object, in which broad low-ionization absorption
is {\em not} seen even though \ion{Mg}{2} is within the spectral
coverage, and H denotes a HiBAL in which the \ion{Mg}{2} region is not
within the spectral coverage or has very low signal-to-noise ratio,
and so whether or not the object is a LoBAL as well as a HiBAL is
unknown.  Thus, all $\CIV$ BAL measurements are coded with `H', all
$\MgII$ BAL measurements are coded with `Lo', and the 86 $1.7\le z\le
2.15$ quasars with both $\CIV$ and $\MgII$ BAL trough measurements
have two entries in the catalog.  There are a handful of $1.7\le z\le
2.15$ LoBALs in Table \ref{tbl:catalog} without an additional table
entry for their $\CIV$ trough; these are all heavily reddened LoBALs
with spectral signal-to-noise ratios too low in the $\CIV$ region to
detect or measure a trough there.  There are no known $\MgII$-only
BALQSOs in this or other BALQSO catalogs.

A significant number of BALQSOs in our catalog have troughs which are
partially resolved individual absorption doublets.  Such systems are
sufficiently different from traditional BALs that some catalog users
may prefer to exclude them, and we have therefore coded them with BAL
subtype 'n'.  However, inspection shows that only a minority may be
true catalog contaminants such as blends of two or more narrow
absorption systems at very similar redshifts or intrinsically narrow
systems included in the catalog due to imperfect fits of the nearby
continuum.  There are 51 such relatively narrow systems in the $\MgII$
catalog (11.2\%) and 1069 such systems in the $\CIV$ catalog (24.6\%).
Further discussion on NALs in BALQSOs is presented by Ganguly
et~al. (2001).

We estimate that our catalog is $95\pm2\%$ complete based on two
independent checks.  First, among catalog LoBALs at $1.7\le z\le
2.15$, $10\pm3$\% were not originally also flagged as HiBALs, and had
to be added manually.  However, proper fitting of the continuum of a
HiBAL (necessary for trough identification) is most difficult for
$z\lesssim 2$ quasars, when the \ion{C}{4} absorption region is close
to the short-wavelength limit of the SDSS spectra, so a 10\% loss rate
is probably an overestimate for the entire catalog.  Second, only 12
of 187 HiBALs ($6.5\pm2.1\%$) from Reichard et al. (2003a) which lie
within our redshift range were not identified by our automated
algorithm.  Seven of those ($3.8\pm1.7\%$) were definite BALQSOs that
we manually added to our catalog, and three of those seven had $z<2$,
where continuum fitting is most difficult.  We adopt the weighted
average of these completeness estimates: $95\pm2\%$.

We now discuss some individual examples of HiBALs, traditional HiBALs,
LoBALs, and FeLoBALs.  The frequencies of BAL phenomena given in the
following subsections are raw only; a future paper will address 
the true frequencies, after correction for selection effects.

\subsection{HiBALs}

Among the 16883 quasars of the SDSS DR3 in the redshift range $1.7 \le
z \le 4.38$, 1756 ($10.4\pm0.2\%$) are identified as traditional
HiBALs, with a nonzero BI in the $\CIV$ region.  This is a $3.5\sigma$
smaller fraction of traditional HiBALs than the $14\pm1\%$ identified
by Reichard et~al. (2003b) at $1.7 \le z \le 4.2$ in the SDSS EDR.  
%z<4.2: 1730/16767 10.3%
%PBH: 2 sigma is not worth speculating on.  %Our method of continuum
%fitting may be one reason for the smaller fraction in our sample: 
%with more template spectra, our continuum fits match the spectra
%better and a few ``borderline'' BALQSOs of low BI in the
%catalog of Reichard et~al. (2003a) are eliminated from our catalog.

\begin{figure}
\plotone{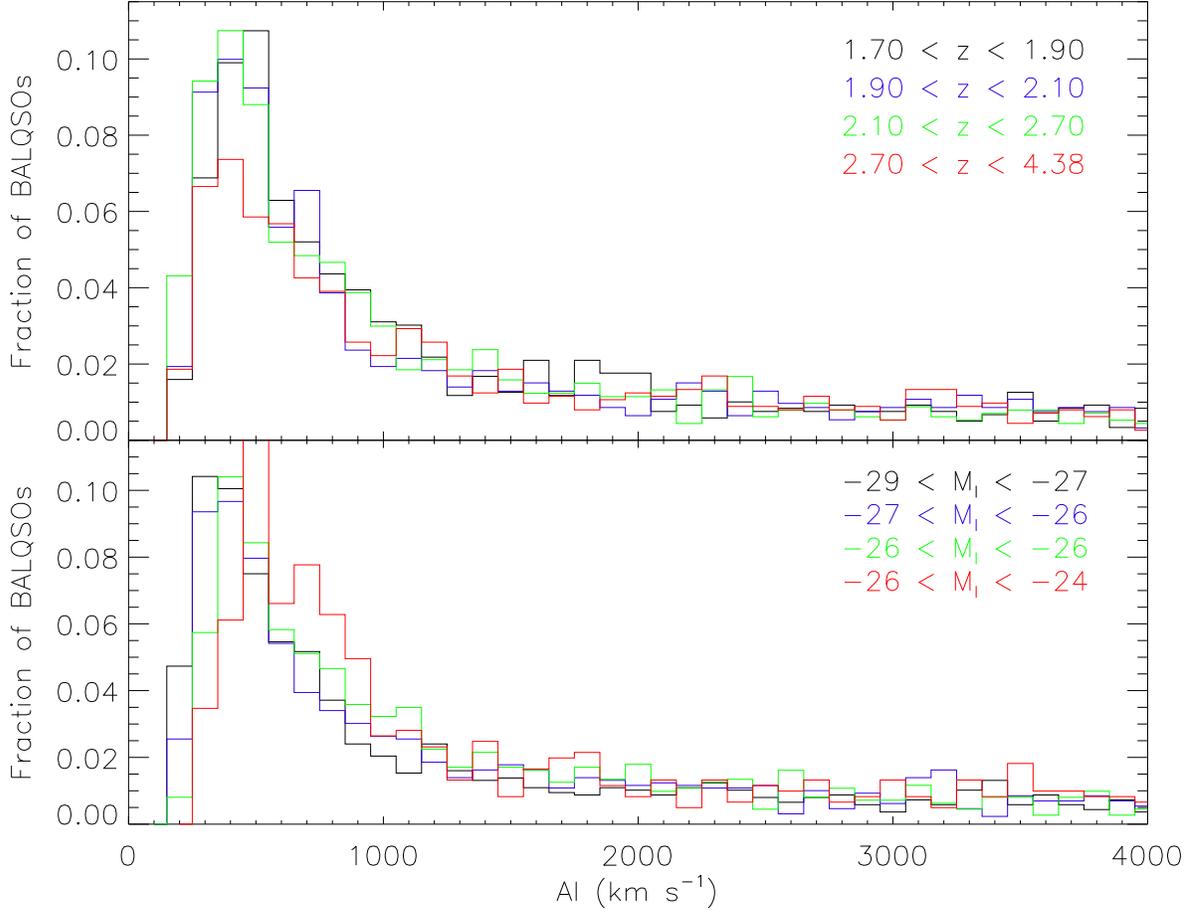}
\caption{The distribution of computed AI values of the 5418 quasars in
  the redshift range $1.7 \le z \le 4.38$ with a nonzero AI (as
  defined in \S 4.3) in the $\CIV$ region.  Each bin is $100~\kms$
  wide.  The peak in the distribution occurs at ${\rm AI} \approx
  400~\kms$.  While there are a substantial number of objects with
  ${\rm AI} \ge 4000~\kms$, they are evenly distributed and there are
  no interesting features in the distribution beyond the displayed
  range. \label{fig:aihist}}
\end{figure}

\begin{figure}
\plotone{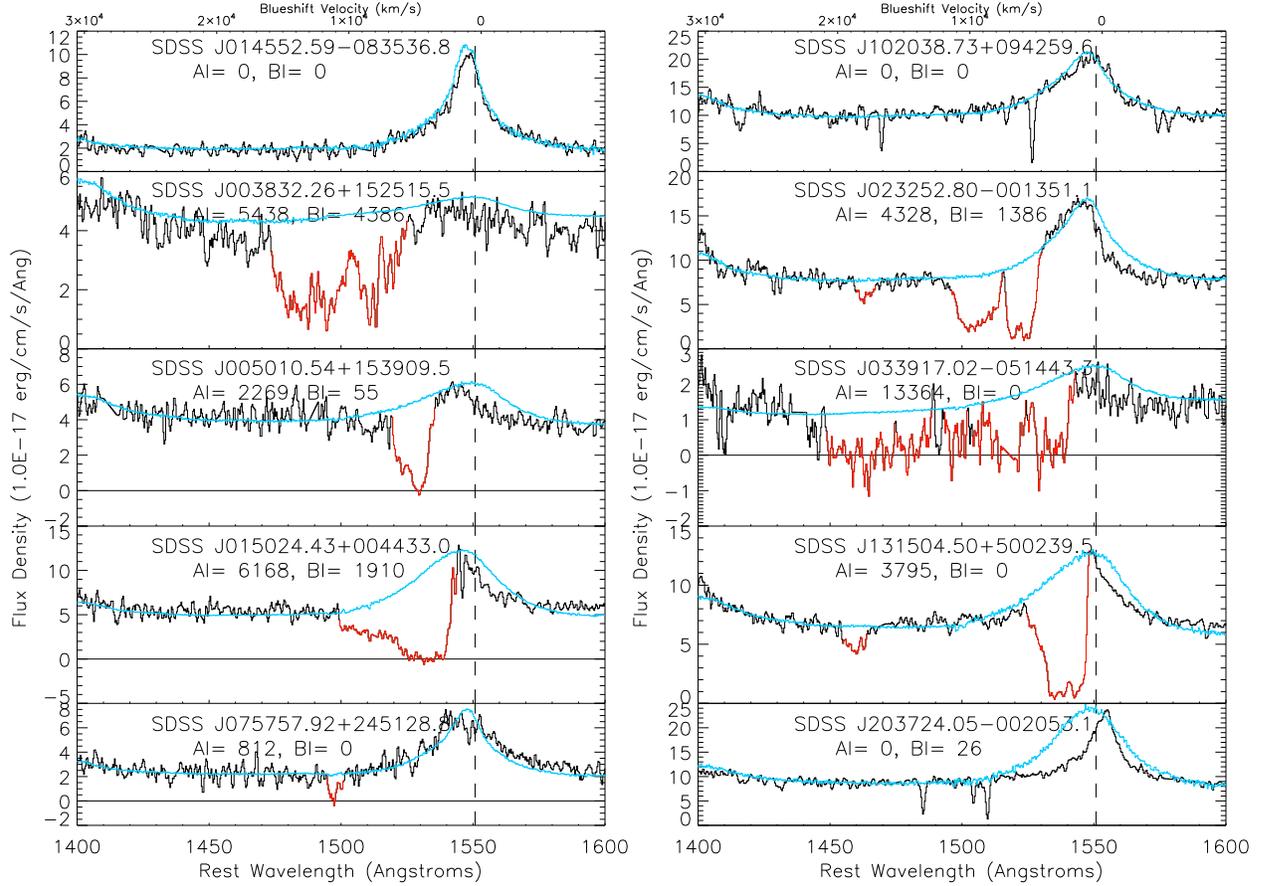}
\caption{Two non-BALQSO spectra, seven example BALQSOs, and one object
  with ${\rm BI}>0$ and ${\rm AI}=0$ plotted in the region 1400-1570
  \AA.  These objects were chosen as examples to show fitted continua
  in the $\CIV$ region and to show the differences between the AI and
  BI.  All ten are discussed individually in \S 5.1.  All spectra are
  smoothed by 3 pixels (roughly the SDSS resolution element).
  Spectral regions identified as BAL troughs (according to the AI
  definition) are designated in red. \label{fig:BALQSO}}
\end{figure}

We identify 4386 HiBALs with ${\rm AI} > 0$ in the $\CIV$ absorption
region.  This amounts to $26.0\pm0.3\%$ of quasars.  The distribution
of AI values among HiBALs is given by Figure \ref{fig:aihist}.  The
observed distribution peaks at ${\rm AI} \approx 400~\kms$, past the
minimum value of ${\rm AI} = 100~\kms$.  However, it does not denote a
peak in the true equivalent width distribution of BAL troughs because
our sample does not include all objects at these AI values: absorption
systems up to 999~\kms\ wide (less than the minimum width for the AI)
can have AI values up to $999~\kms$ and yet are excluded from the
plot.  The AI distribution slowly tails off at large values, with no
new trends for the region not shown (${\rm AI} \ge 4000~\kms$).

Two non-BAL quasars, seven HiBALs, and one quasar with ${\rm BI} > 0$
and ${\rm AI} = 0$ from our catalog are displayed in Figure
\ref{fig:BALQSO}.  All spectra are smoothed by 3 pixels.  We discuss
each object individually.

\begin{enumerate}
     \item SDSS J014552.59$-$083536.8: This object is not a BALQSO,
       but shows the typical continuum fit for a normal $\CIV$ region.

     \item SDSS J102038.73$+$094259.6: This quasar is not a BALQSO.
	The troughs in its spectrum are too narrow to be identified as
	intrinsic broad absorption by our algorithms.

      \item SDSS J003832.26$+$152515.5: The broad troughs in this
	BALQSO make it a traditional BALQSO, but the AI is a better
	description than the BI because the AI includes the absorption
	within the minimum width and depth.  Our line scaling method
	accurately fits the very weak $\CIV$ line.

      \item SDSS J023252.80$-$001351.1: The spectrum of this object
	exhibits two wide and deep troughs near the emission line and
	a shallow and narrower trough around $15000~\kms$ from the
	emission peak.  The BI computation includes only the two
	largest troughs, both of which are narrow enough that the
	inclusion of the minimum width and depth causes the AI to be
	much greater than the BI.  This quasar is also a LoBAL.

      \item SDSS J005010.54$+$153909.5: Although all the broad trough
	in this spectrum is identified in the traditional BALQSO
	definition, the AI is much greater than the BI because the BI
	calculation does not include the minimum width and depth.

      \item SDSS J033917.02$-$051443.3: This quasar's spectrum has low
	signal-to-noise, and we do not require the usual $\chi_{\rm
	trough}^2 > 10$ criterion in order to include the troughs in
	the AI calculation.  The noise prevents any of the troughs
	from being wide enough to be included in the BI calculation.

      \item SDSS J015024.43$+$004433.0: The wide, deep trough in its
	spectrum causes this object to be a traditional BALQSO as well
	as a BALQSO with ${\rm AI}>0$.  However, the BI does not
	include the minimum depth, the minimum width, or the regions
	within $3000~\kms$ of the emission center, and therefore is a
	poorer characterization of the absorption than the AI.  This
	quasar is also a LoBAL.

      \item SDSS J131504.50$+$500239.5: The broad and deep trough in
	this quasar's spectrum is within $3000~\kms$, and the trough
	at $18000~\kms$ is narrower than $2000~\kms$, so it is not
	identified as a traditional BALQSO.  Both of these troughs are
	included in the AI calculation.

      \item SDSS J075757.92$+$245128.8: This quasar spectrum has only
	a relatively narrow trough.  While the trough is unlikely to
	be a contaminant, we designate this BALQSO as an 'n' subtype
	object.  The trough is too narrow for this quasar to be a
	traditional BALQSO.

      \item SDSS J203724.06$-$002053.2: A poor fit to the $\CIV$
	emission line of this non-BALQSO causes the spurious
	identification of a trough from about $3000-5500~\kms$ of the
	emission center.  Our definition of the AI requires troughs
	within $5000~\kms$ to fall at least $2\sigma$ below a linear
	continuum interpolated over the region $\pm 5000~\kms$ of the
	emission center, and therefore does not include this spurious
	trough.  We identify 138 such objects with ${\rm BI} > 0$ and
	${\rm AI} = 0$, all of which are not included in our BALQSO
	catalog because they contain only spurious troughs.

\end{enumerate}

\subsection{LoBALs}

Only 191 non-spurious traditional LoBALs were identified via $\MgII$.
This represents just $0.55\pm0.04\%$ of the 34973 SDSS DR3 quasars in
the redshift range $0.5 \le z \le 2.15$.  The small fraction of LoBALs
found using the traditional BI is not surprising, since $\MgII$ broad
absorption troughs are known to be weaker and narrower than those
found for $\CIV$ (Voit, Weymann, \& Korista 1993).  In addition,
$\MgII$ absorption troughs typically occur nearer the systematic
velocity, within $3000~\kms$ (Reichard et~al. 2003a).  For these
reasons it is more appropriate to discuss our sample of LoBALs
identified through the AI in the $\MgII$ region.

\begin{figure}
\plotone{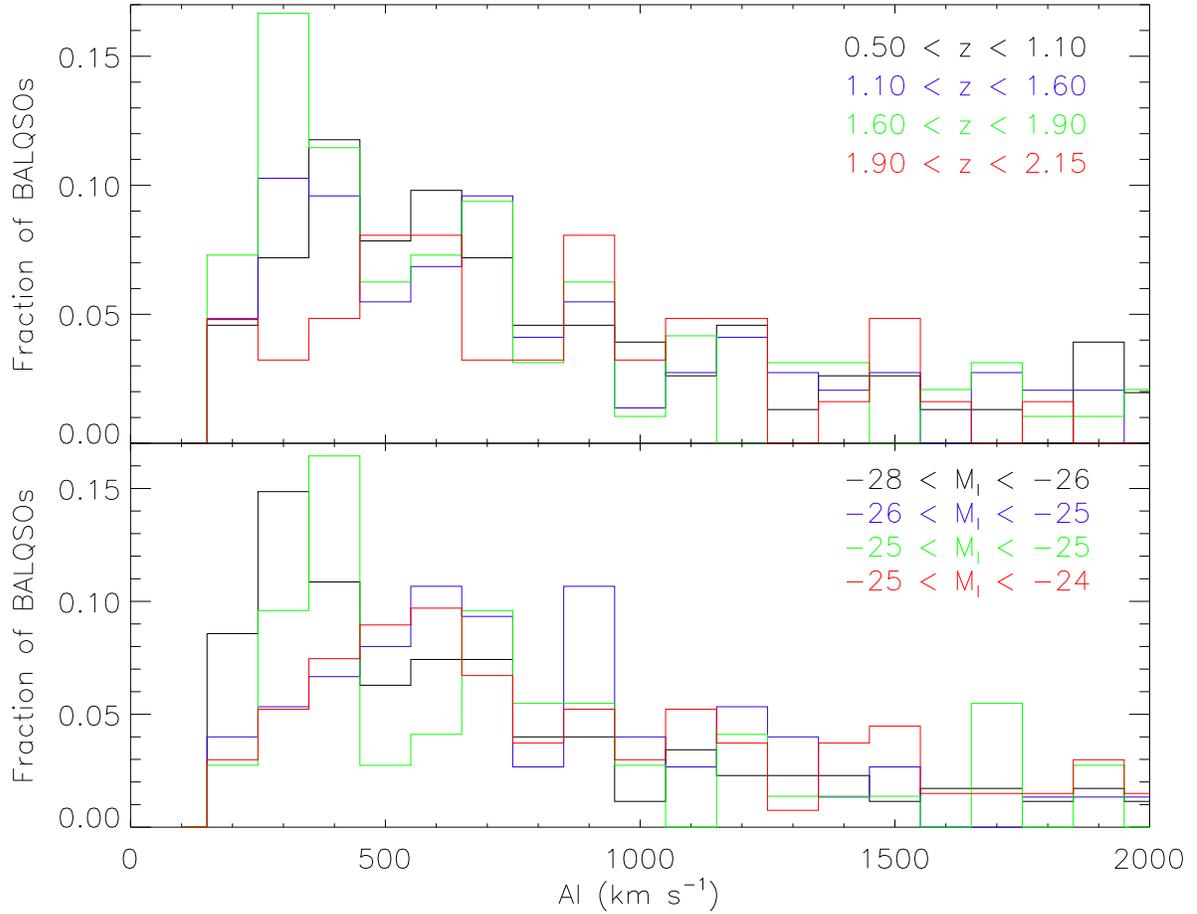}
\caption{The AI distributions of quasars with nonzero $\MgII$ AI in
  different redshift and luminosity bins.  Each bin is $50~\kms$ wide.
%  and the displayed $1\sigma$ error is representative of all redshift
%  or luminosity bins, since all bins were chosen to have approximately
%  the same number of objects.
  The distribution peaks at about ${\rm AI} \approx 350~\kms$, and
  only a few quasars have ${\rm AI} > 2000~\kms$ in the $\MgII$
  region. \label{fig:aihistmgii}}
\end{figure}

\begin{figure}
\plotone{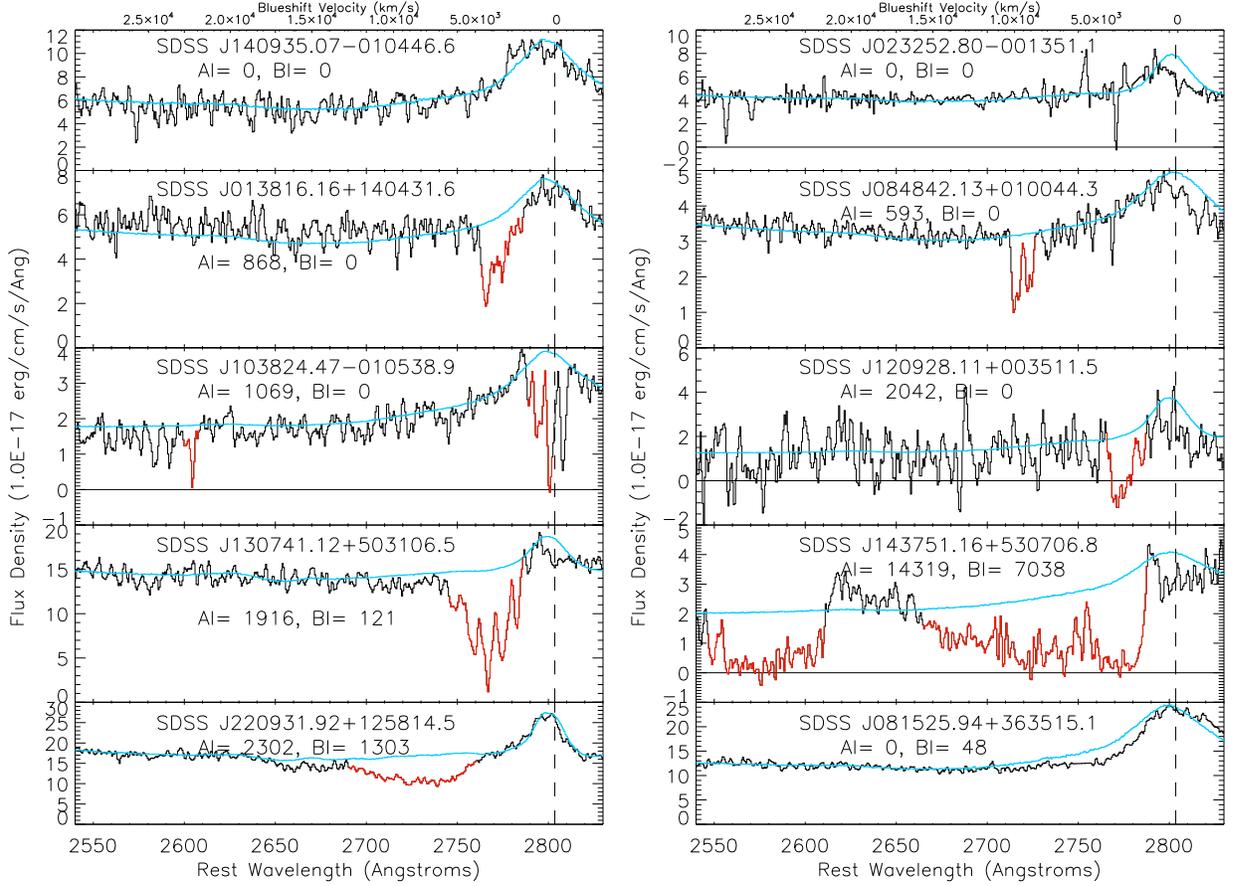}
\caption{Two non-LoBALs, seven LoBAL spectra, and one object with
  ${\rm BI}>0$ and ${\rm AI}=0$ plotted in the region 2570-2830 \AA.
  These objects were chosen as examples to illustrate our definition
  of LoBALs and the differences between the AI and BI.  They are
  discussed individually in \S 5.2.  Spectral regions identified as
  BAL troughs (according to the AI definition) are designated in
  red. \label{fig:LoBAL}}
\end{figure}

We identify 457 $\MgII$ LoBALs, or $1.31\pm0.06\%$ of quasars in our
sample redshift range.  Figure \ref{fig:aihistmgii} gives the
distribution of AI values for these LoBALs.  The distribution peaks at
nearly the same AI value as Figure \ref{fig:aihist} for BALQSOs, but
tails off more swiftly, indicating that $\MgII$ troughs are not
typically as wide and/or deep as $\CIV$ troughs.  This is in agreement
with other studies of LoBALs (e.g., Voit, Weymann, \& Korista 1993).

Two non-LoBALs, seven LoBALs, and one traditional LoBAL with 
$\rm{AI}=0$ are shown in Figure \ref{fig:LoBAL} and discussed
individually below.

\begin{enumerate}
     \item SDSS J140935.07$-$010446.6: This quasar is a non-LoBAL (and
       a non-BALQSO as well), and it is representative of normal
       quasar spectrum and continuum fit in the region 2570-2830 \AA.

     \item SDSS J023252.80$-$001351.1: This object has no absorption
       in the $\MgII$ region and is therefore a non-LoBAL.  However,
       in the $\CIV$ region (shown in Figure \ref{fig:BALQSO}), this
       quasar has ${\rm AI} = 4328~\kms$ and is therefore a BALQSO.
       This quasar demonstrates that not all BALQSOs are LoBALs.

     \item SDSS J013816.16$+$140431.6: This quasar spectrum exhibits a
       typical $\MgII$ absorption trough.  When compared to typical
       $\CIV$ absorption troughs, most LoBALs (and this object in
       particular) exhibit absorption which is shallower, narrower,
       and nearer the emission region.  The trough is too near the
       emission to be identified in the BI calculation.

     \item SDSS J084842.13$+$010044.3: The troughs in this quasar
       spectrum are too narrow (even joined as they are) to be
       identified in the BI calculation, but they are included in the
       AI calculation.

     \item SDSS J103824.47$-$010538.9: The spectrum of this quasar
       features several deep, narrow troughs.  Although the troughs
       are unlikely to be caused by intervening systems, we designate
       this objects as a 'n' subtype LOBAL.  None of these troughs is
       sufficiently wide enough for a BI-based identification.

     \item SDSS J120928.11$+$003511.5: Despite the low signal-to-noise
       ratio of this quasar spectrum, a broad absorption trough is
       identified near the emission region.  This trough is too
       shallow for identification by the BI.
  
     \item SDSS J130741.12$+$503106.5: This quasar is one of the few
       LoBALs with ${\rm BI}>0$ as well as ${\rm AI}>0$.  However, of
       its obvious broad absorption trough, only the small part bluer
       than $3000~\kms$ of the emission line is included by the BI
       calculation.  The AI provides a much better metric of the
       absorption than the BI.

     \item SDSS J143751.16$+$530706.8: This quasar spectrum exhibits
       two broad, deep troughs.  Although the quasar is identified as
       a traditional BALQSO, the AI includes the minimum depth and
       width of each trough in its calculation, and better
       characterizes the absorption than the BI.  This quasar is also
       a FeLoBAL.

     \item SDSS J220931.92$+$125814.5: The absorption trough in the
       quasar spectrum is broad, but rather shallow compared to the
       other LoBALs shown here.  Since the AI includes the minimum
       depth and width in its calculation, it provides a much better
       description of the absorption in this quasar than the BI.

     \item SDSS J081525.94$+$363515.1: This non-BALQSO has ${\rm
       BI}>0$ but ${\rm AI}=0$.  The nonzero BI identifies a spurious
       trough from about $3000-6000~\kms$ of the emission center.
       This region is below the fitted continuum only because of the
       inadequate fit to an asymmetric emission profile.  The AI
       definition, however, requires that absorption troughs within
       $5000~\kms$ of the emission center fall at least $2\sigma$
       below a linear continuum interpolated over the region $\pm
       5000~\kms$.  We identify 137 ${\rm BI}>{\rm AI}$ objects in the
       $\MgII$ region, all of which have only spurious troughs and are
       not included in our BALQSO catalog.

\end{enumerate}

\subsection{FeLoBALs}

\begin{figure}
\plotone{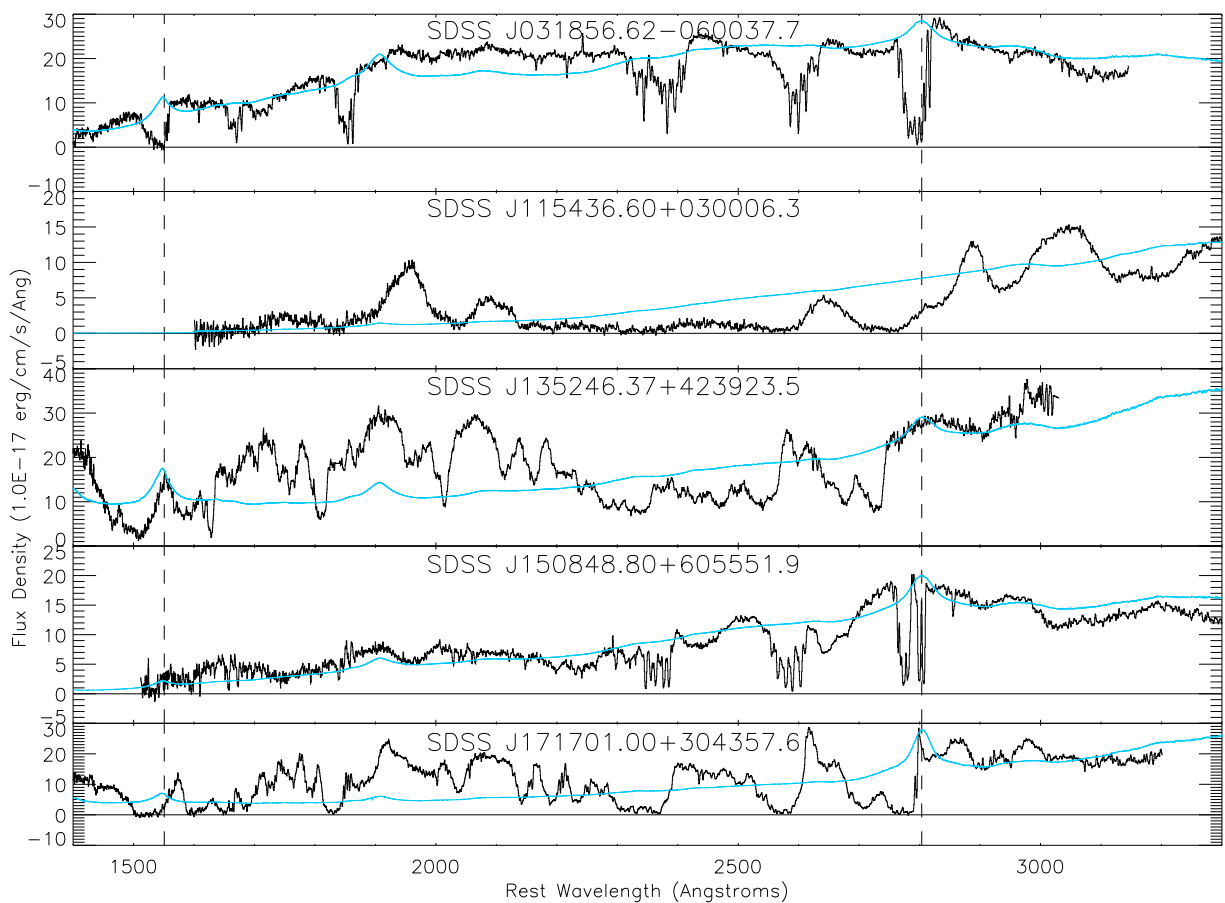}
\caption{Five FeLoBALs at rest frame wavelengths 1400-3300 \AA, with
  the centers of $\CIV$ and $\MgII$ emission shown as vertical dashed
  lines.  Overplotted in blue on each spectrum is the $\MgII$
  continuum fit (quasars with $1.7 \le z \le 2.15$ also have a
  separate $\CIV$ continuum fit).  All FeLoBALs in our catalog were
  identified by visual inspection.  All five of these FeLoBALs exhibit
  strong absorption throughout their spectra, making continuum fitting
  and automatic identification extremely
  challenging. \label{fig:FeLoBALs}}
\end{figure}

We searched for FeLoBALs by visual inspection of our BALQSO sample.
We identified 138 unique confirmed and candidate FeLoBALs, one of
which (SDSS J094317.60$+$541705.1) was added to the catalog by hand
since its spectrum shortward of $\MgII$ is completely obliterated by
overlapping troughs.  Up to $0.33\pm0.03\%$ of SDSS quasars at $0.5\le
z\le 4.38$ are FeLoBALs.  Five FeLoBALs are shown in Figure
\ref{fig:FeLoBALs} and are discussed below.

\begin{enumerate}
     \item SDSS J031856.62$-$060037.7: The spectrum of this FeLoBAL
       exhibits several deep troughs associated with $\CIV$,
       $\AlIII$, $\MgII$, $\CIII$, $\FeII$, and $\FeIII$.  Although
       the AI in the $\MgII$ region is likely accurate, the widespread
       absorption causes the continuum to be slightly understimated
       outside the $\MgII$ region, since absorption is only weighted
       out of the fitting in the $\MgII$ region.  This quasar is
       discussed in more detail by Hall et~al. (2002).

     \item SDSS J115436.60$+$030006.3: The absorption troughs in the
       spectrum of this FeLoBAL are extremely broad and overlap, with
       only a small fraction of the spectrum remaining unabsorbed.
       The $\MgII$ emission is almost invisible due to absorption from
       $\MgI$ $\lambda$2853\,\AA\ absorption.  The continuum is
       greatly underestimated in the fitting.  Further discussion for
       this quasar can be found in Hall et~al. (2002).

     \item SDSS J135246.37$+$423923.5: The absorption troughs in the
       spectrum of this quasar are very broad, although they are not
       quite so deep or overlapping as those of SDSS J1154+0300.  Many
       of the troughs also appear more blueshifted, so that the
       $\CIV$, $\CIII$, and $\MgII$ emission is more apparent than in
       the other FeLoBALs shown here.

     \item SDSS J150848.80$+$605551.9: With several deep and
       relatively narrow troughs, this quasar appears to be a more
       heavily reddened version of the FeLoBAL SDSS J0318$-$0600.  As
       in SDSS J0318$-$0600, the continuum is only slightly
       underestimated in the fit to this FeLoBAL.

     \item SDSS J171701.00$+$304357.6: The spectrum of this FeLoBAL
       exhibits broad, deep troughs throughout the spectrum.  The
       continuum fit is very poor, and the absorption in both the
       $\CIV$ and the $\MgII$ regions are probably underestimated.
\end{enumerate}

\subsection{Manually Included Objects}

\begin{figure}
\plotone{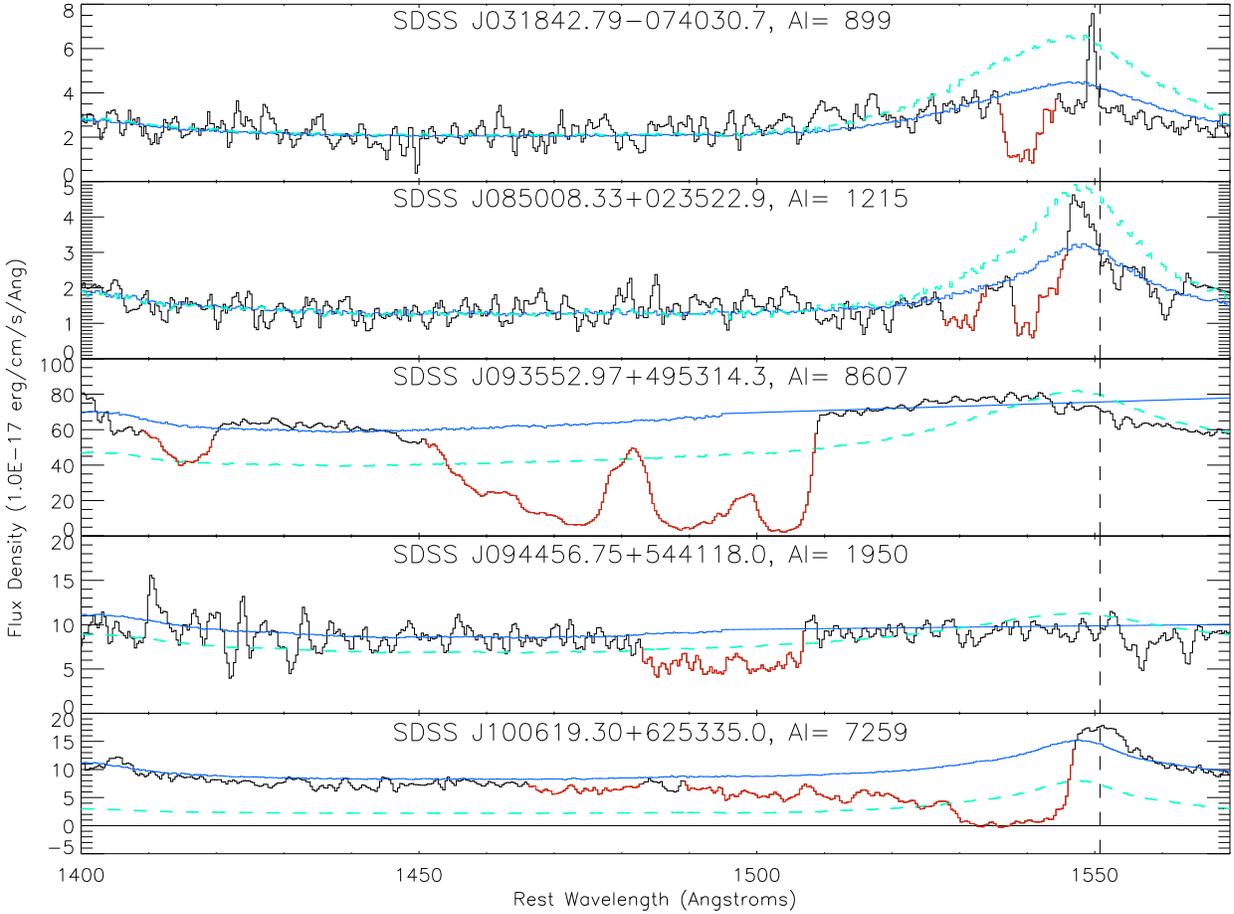}
\caption{Five quasars with manually adjusted best-fit templates.  The
  original automatically determined best-fit template is shown as
  the dashed teal line and the manually adjusted best-fit template
  is shown as the blue line.  Each spectrum and its manually adjusted
  template is described in \S 5.4. \label{fig:adjusted}}
\end{figure}

We visually inspect the best template fits for all quasars, and adjust
the best-fit templates for 347 quasars in the $\CIV$ region and 3
quasars in the $\MgII$ region.  In some cases the quasar has an unusual
emission line profile and the automatically scaled emission line of
the best-fit template is manually adjusted to prevent over- or
under-estimation of the AI.  In other objects, the normalization
region of the quasar spectrum (used to determine the scale factor) is
especially noisy, and the continuum scale must be adjusted.  
In Figure \ref{fig:adjusted} we display five
quasars in the $\CIV$ region with their automatic best-fit templates
and their manually adjusted best-fit templates.  Each object shown in
Figure \ref{fig:adjusted} is individually discussed below.

\begin{enumerate}
     \item 031842.79$-$074030.7: The $\CIV$ emission region is
       contaminated by narrow H$\beta$ emission from an emission-line
       galaxy along the line of sight, which causes the automatically
       determined best-fit template to over-estimate the AI.  We
       manually reduce the line scaling of the best-fit template to
       accurately determine the AI of this BALQSO.

     \item 085008.33$+$023522.9: The automically scaled emission line
       region does not accurately represent the $\CIV$ line profile of
       this quasar.  We adjust the line scaling and more accurately
       determine the AI.

     \item 093552.97$+$495314.3: Widespread broad absorption causes the
       best-fit template continuum to be underestimated between the
       $\SiIV$ and $\CIV$ emission lines.  We increase the scaling
       factor in order to measure all of the absorption.

     \item 094456.75$+$544118.0: While the absorption is not so
       widespread as in the previous object, the best-fit template
       continuum is still underestimated and this object is falsely
       determined to be a non-BALQSO.  After manually adjusting the
       scaling factor, this quasar is correctly classified as a BAL
       with a shallow trough.

     \item 100619.30$+$625335.0: This is one of the few objects for
       which the automatic template fitting failed for no obvious
       reason.  The manually adjusted template provides an accurate
       continuum from which to measure the AI.

\end{enumerate}

\section{Discussion: Spectral Properties of BALQSOs}

The size of our catalog presents the opportunity for a statistical
study of the nature of BALQSOs.  In this paper we study what can be
learned directly from the results of the fitting process; in Hall et
al. (2006, in preparation) we address issues that require additional
analysis.

\begin{figure}
\plotone{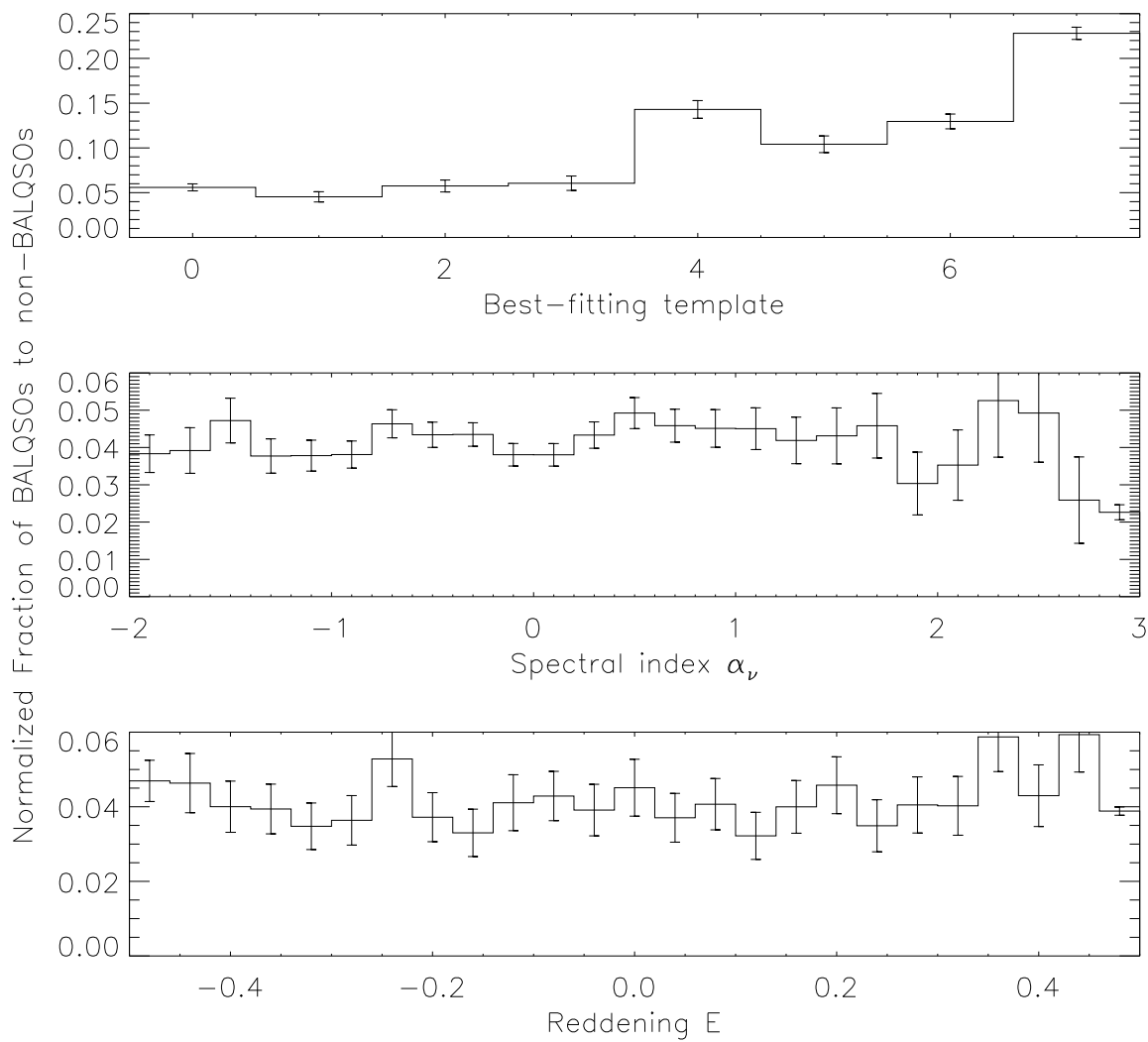}
\caption{Histograms of the normalized ratio of $\CIV$ BALQSOs to
  non-BALQSOs with best template fit, spectral index, and reddening.
  Templates 0-3 are in the less luminous bin and 3 and 7 are the
  widest line width bins.  Higher values of the spectral index
  represent bluer continua, while higher values of the reddening
  represent redder continua.  We include only BALQSOs with spectral
  SNR$>9$ in order to remove any luminosity effects on the fitting,
  since BALQSOs with low signal-to-noise may only be identified if
  they are more luminous.  The preference for BALQSOs to be fit by the
  more luminous and widest line width template is probably a physical
  effect. \label{fig:spechist}}
\end{figure}

\begin{figure}
\plotone{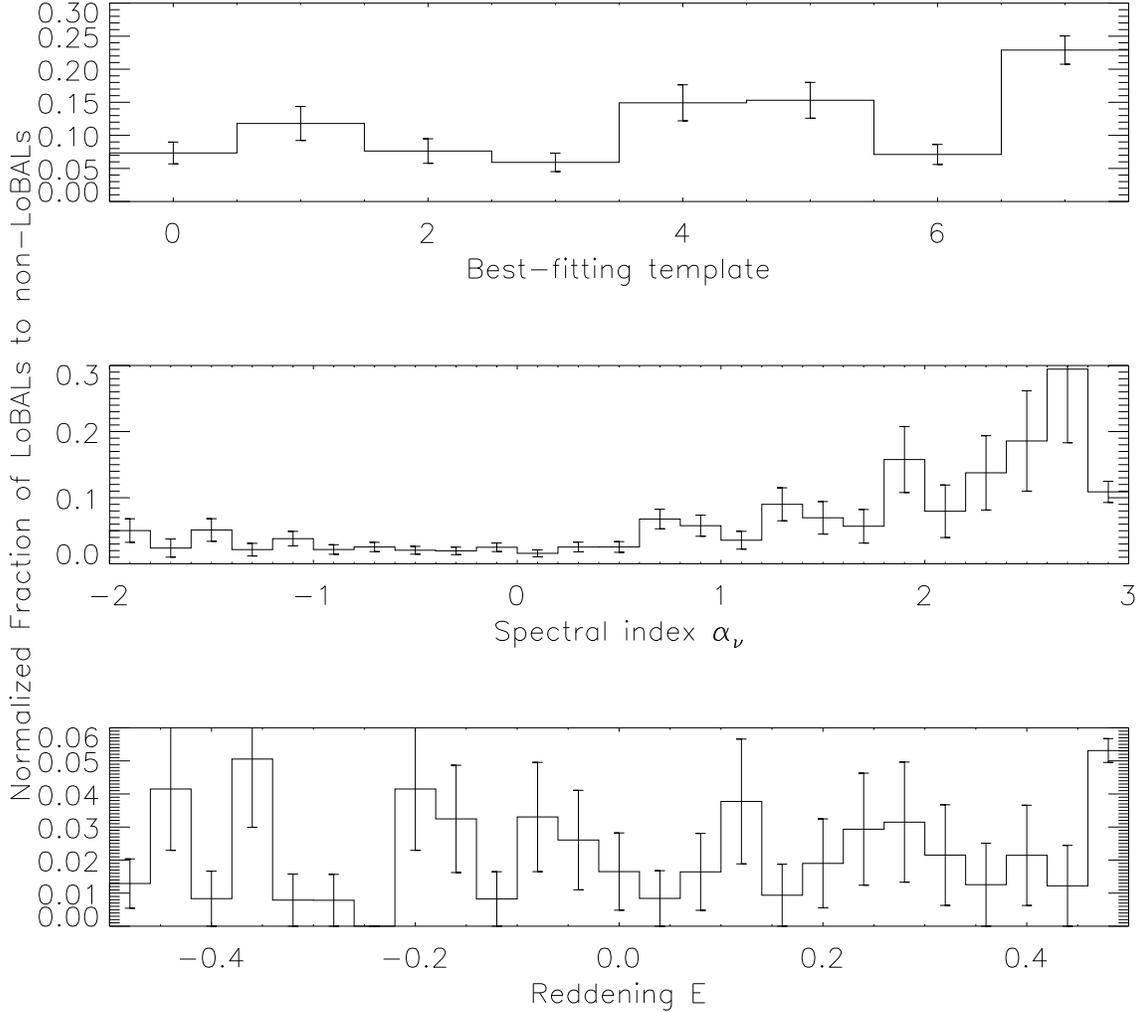}
\caption{Histograms of the normalized ratio of $\MgII$ LoBALs to
  non-LoBALs with best template fit, spectral index, and reddening.
  Templates 0-3 are in the less luminous bin and 3 and 7 are the
  widest line width bins.  We include only LoBALs with spectral
  SNR$>9$ in order to remove any luminosity effects on the fitting.
  As with BALQSOs, the preference to be fit by the more luminous and
  widest line width template is probably a physical effect.  The
  preference for fitting by bluer spectral indices may be a bias since
  fitting the possibly highly absorbed LoBAL continua may prefer more
  extreme fits. \label{fig:spechistmgii}}
\end{figure}

We investigate the spectral properties of the BALQSOs in our sample by
comparing their properties to those of the non-BALQSOs in our parent
data set.  Figure \ref{fig:spechist} shows the distribution of BALQSOs
with the best-fitting template, spectral index, and reddening (the
latter two parameters are highly degenerate and not necessarily
physical).  Figure \ref{fig:spechistmgii} shows the same distributions
of LoBALs.  We include only BALQSOs and LoBALs of spectral SNR$>9$
because the low signal-to-noise BALQSOs and LoBALs that we identify
may be preferentially more luminous, with correspondingly broader
emission lines.  Figures \ref{fig:spechist} and \ref{fig:spechistmgii}
should therefore have no artificial luminosity effects caused by our
fitting constraints.  The top histograms indicate that BALQSOs and
LoBALs favor a continuum fit by the more luminous and widest line
width template.  We interpret this as a fit to wider line width
templates in general: since line width is correlated with luminosity,
less luminous BALQSOs and LoBALs with wider line widths may prefer a
fit by more luminous templates instead of the less luminous widest
line width template.  This preference suggests that BALQSOs and LoBALs
have broader emission lines than other quasars.  If the emission-line
width depends in part on the orientation of a disk-like emission
region to our line of sight (Murray \& Chiang 1998; Peterson 2003),
this preference suggests that broad absorption troughs are
preferentially observed along low-latitude sightlines above the disk.
This preference is also consistent with previous work: Richards et~al.
(2002) showed that quasars with large systematic emission line
blueshifts tend to have broader lines, and Reichard et~al. (2003b)
showed that BALQSOs tend to have large systematic blueshifts.

\begin{figure}
\plotone{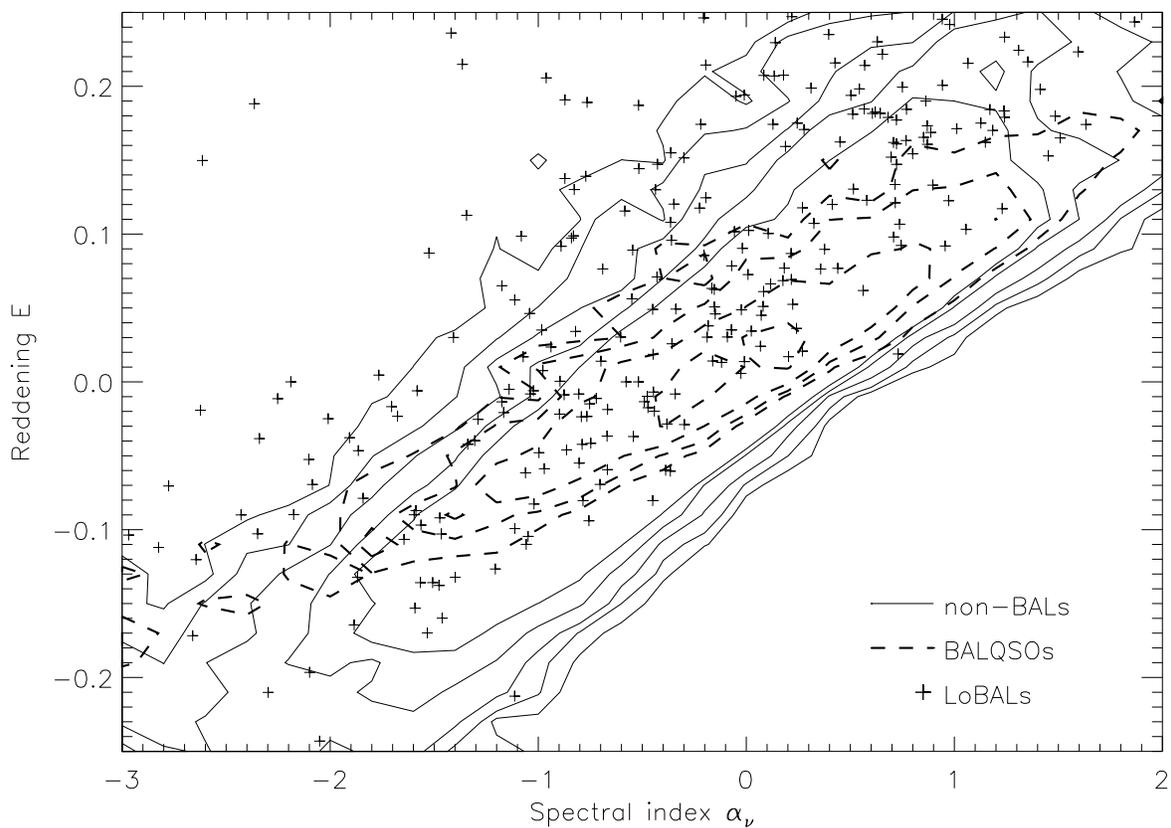}
\caption{Reddening versus spectral index for BALQSOs and non-BALQSOs.
The solid contours show the density of non-BALQSOs, the dashed
contours show the density of BALQSOs, and LoBALs are overplotted as
symbols.  The spectral index and reddening are highly degenerate over
the limited wavelength baseline of the SDSS spectra, so the formal
'best-fit' values are often unphysical.  Despite this degeneracy, all
three distributions are statistically different from each other at
high significance, in the sense that BALQSOs (and especially LoBALs)
are more reddened than non-BALQSOs.
\label{fig:Evsa}}
\end{figure}

Since the spectral index and reddening are degenerate in our fitting
algorithms, we study the properties of both with regards to BALQSOs,
LoBALs, and non-BALQSOs through Figure \ref{fig:Evsa}.  While most
LoBALs and BALQSOs are located within the contours of the non-BALQSO
population, a large fraction of BALQSOs are shifted to the upper left
of the plot, indicating that BALQSOs on average are more intrinsically
reddened than non-BALQSOs similar to the result of Reichard et~al.
(2003b).  A two-dimensional Kolmogorov-Smirnov (K-S) statistical test
(Press et~al. 1998) on the spectral index and reddening distribution
reveals probabilities of less than $10^{-7}$ that BALQSOs and
non-BALQSOs share the same spectral index and reddening distribution,
or that LoBALs and non-BALQSOs share the same distribution, or that
LoBALs and BALQSOs share the same distribution.  The shift of BALQSOs
in spectral index-reddening space creates a selection effect that must
be corrected in order to produce a true fraction of BALQSOs.

\section{Conclusions}

\begin{enumerate}
     \item Our definition of broad absorption allows for the
	identification and characterization of absorption troughs near
	the emission line peak and as narrow as $1000~\kms$.  The AI
	provides a more complete description of broad absorption than
	the traditional BI.

     \item BALQSOs defined by ${\rm AI}>0$ comprise approximately 26\%
        of quasars, $\MgII$ LoBALs about 1.3\%, and FeLoBALs about
        0.3\%.  These are raw fractions only, and do not take into
        account the larger average reddening seen in BALQSOs as
        compared to non-BALQSOs.

     \item BALQSOs appear to have wider emission lines than
	non-BALQSOs, on average.  This supports an AGN model in which
	the broad emission and absorption regions are in a disk-like
	configuration.
\end{enumerate}

%% In future studies of BALQSOs in the SDSS DR3, we plan these major
%% improvements:
%% 
%% \begin{itemize}
%%      \item We will calculate the AI for other lines (i.e. $\AlIII$),
%%        probing the intermediate ionization lines between LoBALs and
%%        HiBALs.
%% 
%%      \item The selection effects should be studied and applied in
%%        more detail to produce a true fraction of BALQSOs.  This
%%        could provide constraints on the geometry of quasar intrinsic
%%        absorption systems.
%% 
%%      \item The correlation between emission line width and the
%%        presence of BALs should be further investigated.  More
%%        accurate emission line profile fitting will provide more data
%%        with which to study this apparent correlation.
%% 
%%      \item Further analysis of the properties of this sample of
%%        BALQSOs will appear in Hall et~al. 2005.
%% 
%% \end{itemize}

\acknowledgments

We thank the referee for many helpful comments.  This research was
partially supported by grant NSF AST03-07582 (JRT, DPS, DVB).  JRT
acknowledges support from a Schreyer's Honors College Summer Research
Scholarship.  PBH acknowledges support from a NSERC Discovery Grant.
GTR was supported in part by a Gordon and Betty Moore Fellowship in
Data Intensive Sciences.

Funding for the creation and distribution of the SDSS Archive has been
provided by the Alfred P. Sloan Foundation, the Participating
Institutions, the National Aeronautics and Space Administration, the
National Science Foundation, the U.S. Department of Energy, the
Japanese Monbukagakusho, and the Max Planck Society. The SDSS Web site
is http://www.sdss.org/.

The SDSS is managed by the Astrophysical Research Consortium (ARC) for
the Participating Institutions. The Participating Institutions are The
University of Chicago, Fermilab, the Institute for Advanced Study, the
Japan Participation Group, The Johns Hopkins University, the Korean
Scientist Group, Los Alamos National Laboratory, the
Max-Planck-Institute for Astronomy (MPIA), the Max-Planck-Institute
for Astrophysics (MPA), New Mexico State University, University of
Pittsburgh, University of Portsmouth, Princeton University, the United
States Naval Observatory, and the University of Washington.

\appendix
\section{Supplementary list of visually identified HiBALs at $z>4.38$}
The upper limit redshift of our BAL catalog is $z=4.38$.  BALQSOs can
be identified at higher redshifts in SDSS spectra, but calculating
their AI or BI values is difficult since normalization and continuum
fitting of high-redshift spectra can be problematic.  Nonetheless, to
further the study of BALQSOs at all redshifts we provide in Table
\ref{tbl:hiz} a list of 32 BALQSOs and 7 BALQSO candidates visually
identified from the 213 $z>4.38$ quasars in the SDSS DR3 quasar
catalog.  Only one, or possibly two, can be identified as LoBALs from
the SDSS spectra alone.

The raw fraction of HiBALs at $z>4.38$ is $15.0^{+2.9}_{-2.5}$\%, 
higher than but consistent with the fraction at lower redshifts.  
However, SDSS spectra of $z>4.38$ quasars are generally of 
lower signal-to-noise ratio than SDSS spectra of lower-redshift HiBALs
(average $i$-band signal-to-noise ratios $5.7\pm2.8$ vs. $9.9\pm5.6$,
respectively).  Therefore, weak BAL troughs will be {\em
underrepresented} in our $z>4.38$ sample and the true incidence of
HiBALs in it will be even larger, perhaps indicating a real increase
in the HiBAL fraction at high redshift.  Careful study of even larger
samples of BALQSOs will be needed to put this possibility on firm
statistical footing.

\clearpage

\begin{deluxetable}{cccr}
\tablecolumns{4}
\tablewidth{250pt}
\tablecaption{Quasar Template Properties \label{tbl:templates}}
\tablehead{
\colhead{Composite} & \colhead{Fitting} & \colhead{Region} & \colhead{Avg. \# per} \\
\colhead{$z_{\rm min}-z_{\rm max}$} & \colhead{$z_{\rm min}-z_{\rm max}$} & \colhead{to fit} & \colhead{template} }
\startdata
2.30-4.90 & 2.30-4.38 & $\CIV$ & 56 \\
1.90-2.30 & 1.90-2.30 & $\CIV$ & 410 \\
1.52-1.90 & 1.70-1.90 & $\CIV$ & 372 \\
1.40-2.20 & 1.40-2.15 & $\MgII$ & 1576 \\
0.90-1.40 & 0.90-1.40 & $\MgII$ & 1552 \\
0.39-0.90 & 0.50-0.90 & $\MgII$ & 1279 \\
\enddata
\end{deluxetable}

\begin{deluxetable}{rrrl}
\tablecolumns{4}
\tablewidth{300pt}
\tablecaption{Wavelength Region Weights \label{tbl:weights}}
\tablehead{
\colhead{$\lambda_{\rm i}$ (\AA)} & \colhead{$\lambda_{\rm f}$ (\AA)} & \colhead{Weight} & \colhead{Description}}
\startdata
0 & 1055 & 0.0 & $\Lyb$ and bluer \\
1055 & 1160 & 0.0 & $\Lya$ forest \\
1160 & 1240 & 0.0 & $\Lya$ emission, $\NV$ blue wing \\
1240 & 1290 & 0.5 & $\NV$ red wing \\
1290 & 1360 & 1.0\tablenotemark{a} & $\SiIV$ absorption region \\
1360 & 1402 & 0.0 & $\SiIV$/$\OIV$ blue wing \\
1402 & 1446 & 0.5 & $\SiIV$/$\OIV$ red wing \\
1425 & 1524 & 1.0\tablenotemark{b} & $\CIV$ absorption region \\
1524 & 1577 & 1.0\tablenotemark{b} & $\CIV$ emission \\
1750 & 1830 & 0.5 & $\AlIII$ absorption region \\
1830 & 1857 & 0.0 & $\CIII$ $\FeIII$ blue wing \\
1857 & 1976 & 0.5 & $\CIII$ $\FeIII$ red wing \\
2579 & 2757 & 1.0\tablenotemark{c} & $\MgII$ absorption region \\
2757 & 2850 & 1.0\tablenotemark{c} & $\MgII$ emission \\
3394 & 3446 & 0.0 & $\NeV$ emission \\
3714 & 3740 & 0.0 & $\OII$ emission \\
3850 & 3884 & 0.0 & $\NeIII$ emission \\
4050 & 9200 & 0.0 & $\Hd$, beyond power-law \\
\multicolumn{2}{c}{$\sigma(\lambda)=0$} & 0.0 & Interpolated pixels \\
\enddata
\tablenotetext{a}{Velocities in the $\SiIV$ region corresponding to
regions of flux identified with absorption in the $\CIV$ region were
assigned $w(\lambda)=0$.}
\tablenotetext{b}{Regions of flux below the first-pass fit by
$4\sigma$ or more were assigned $w(\lambda)=0$.}
\tablenotetext{c}{Regions of flux $2\sigma$ or more below a linear
continuum from $\pm 5000~\kms$ of the emission line in the first-pass
fit were assigned $w(\lambda)=0$, and regions above this threshold
were assigned $w(\lambda)=0.5$.}
\end{deluxetable}

\begin{deluxetable}{rrl}
\tablecolumns{3}
\tablecaption{SDSS DR3 BALQSO Electronic Catalog Format \label{tbl:catdesc}}
\tablehead{
  \colhead{Column} & 
  \colhead{Format} & 
  \colhead{Description}
}
\startdata
1 & A18 & SDSS DR3 Designation hhmmss.ss+ddmmss.s (J2000) \\
2 & I4 & Spectroscopic Plate Number \\
3 & I5 & Modified Julian Date of spectroscopic observation \\
4 & I3 & Spectroscopic Fiber Number \\
5 & F5.3 & Redshift \\
6 & F5.2 & PSF $i$ magnitude (not corrected for Galactic extinction) \\
7 & F6.2 & Absolute $i$ magnitude \\
8 & A4 & Subtype\tablenotemark{a}: H for HiBAL sample, Lo for LoBAL sample, F for FeLoBAL \\
%8 & A4 & Type: MgII for LoBAL, CIV for HiBAL \\
9 & I6 & Absorption Index, AI (km s$^{-1}$) \\
10 & F6.2 & Error in the Absorption Index, $\sigma_{\rm AI}$ (km s$^{-1}$) \\
11 & I6 & Balnicity Index, BI (km s$^{-1}$) \\
12 & F6.2 & Error in the Balnicity Index, $\sigma_{\rm BI}$ (km s$^{-1}$) \\
13 & I1 & Best-fit template number \\
14 & F6.2 & Spectral index $\alpha_\nu$ used by the best-fit template \\
15 & F6.3 & Reddening $E(B-V)$ used by the best-fit template \\
16 & F7.2 & $\chi^2_0$ value of the best fit \\
17 & I2 & Number of distinct BAL troughs \\
18 & I5 & Widest BAL trough (km s$^{-1}$) \\
19 & F4.2 & Deepest part of any BAL trough \\
20 & I6 & Weighted average velocity of the BAL troughs (km s$^{-1}$) \\
21 & I6 & Maximum velocity of the BAL troughs from the emission line (km s$^{-1}$) \\
22 & F4.1 & SDSS spectrum signal-to-noise ratio in the $i$-band \\
\enddata
\tablenotetext{a}{For full details on subtype codes, see \S 5 and Table 4.}
\end{deluxetable}

\clearpage
\begin{deluxetable}{lcrrrcrrcrrrrrrrr}
\rotate
\tablecaption{SDSS DR3 BALQSO catalog (page 1) \label{tbl:catalog}}
\tablecolumns{17}
\tablewidth{620pt}
\tabletypesize{\scriptsize}
\tablehead{
  \colhead{Quasar (SDSS J)} &
  \colhead{Spectrum\tablenotemark{a}} &
  \colhead{$z$} &
  \colhead{$i$} &
  \colhead{$M_i$} &
  \colhead{Type\tablenotemark{b}} &
  \colhead{AI} &
  %\colhead{$\sigma_{\rm AI}$} &
  \colhead{BI} &
  %\colhead{$\sigma_{\rm BI}$} &
  \colhead{T\tablenotemark{c}} &
  \colhead{$\alpha_{\nu}$} &
  \colhead{$E$$($$B$$-$$V$$)$} &
  \colhead{$\chi_{0,\rm{fit}}^2$} &
  \colhead{$n$\tablenotemark{d}} &
  \colhead{widest} &
  \colhead{deepest} &
  \colhead{$v_{\rm wtavg}$} &
  \colhead{$v_{\rm max}$} }
\startdata
000009.26$+$151754.5 &  751-52251-354 & 1.199 & 19.17 & $-25.08$ &   Lo &   1693 &      0 & 3 &  -0.97 & -0.059 &    1.27 &  1 &  4141 & 0.82 &   2217 &   4797 \\
000056.89$-$010409.7 &  387-51791-098 & 2.106 & 19.28 & $-26.29$ &   Hi &   4733 &   1076 & 0 &   0.61 &  0.160 &    1.93 &  1 &  7453 & 1.00 &   4439 &   8639 \\
000103.85$-$104630.3 &  650-52143-133 & 2.081 & 18.15 & $-27.38$ &   Hi &   1222 &      0 & 4 &  -0.36 &  0.022 &    2.17 &  1 &  3451 & 0.66 &   2619 &   4287 \\
000119.64$+$154828.8 &  750-52235-566 & 1.924 & 19.02 & $-26.34$ &   HL &   7299 &   5402 & 0 &   0.44 &  0.091 &    1.61 &  1 & 12067 & 0.91 &   9291 &  15417 \\
000119.64$+$154828.8 &  750-52235-566 & 1.924 & 19.02 & $-26.34$ &   Lo &    502 &      0 & 0 &   0.41 &  0.120 &    2.43 &  2 &  1172 & 0.36 &  10321 &  11737 \\
000303.35$-$105150.6 &  650-52143-048 & 3.647 & 19.25 & $-27.53$ &    H &   1808 &      0 & 0 &  -7.13 & -0.450 &    3.42 &  2 &  2277 & 0.91 &   5588 &  14961 \\
000335.20$+$144743.6 &  750-52235-036 & 3.484 & 20.07 & $-26.67$ &   nH &    758 &      0 & 7 &   2.95 &  0.246 &    1.25 &  1 &  1518 & 0.79 &    859 &   1603 \\
000913.76$-$095754.5 &  651-52141-519 & 2.069 & 18.68 & $-26.85$ &   Hi &   3724 &    852 & 7 &  -0.70 & -0.018 &    1.44 &  1 &  8005 & 0.71 &   4631 &   8982 \\
001025.90$+$005447.6 &  389-51795-332 & 2.847 & 18.94 & $-27.26$ &    H &   4787 &   2557 & 1 &  -1.61 & -0.076 &    4.15 &  2 &  5241 & 0.88 &  11439 &  17378 \\
001053.56$+$000642.9 &  389-51795-348 & 1.879 & 18.98 & $-26.30$ &  nHi &    454 &      0 & 6 &  -0.49 &  0.006 &    1.46 &  1 &  1864 & 0.57 &   2245 &   3326 \\
001130.55$+$005550.7 &  389-51795-339 & 2.309 & 18.34 & $-27.41$ &   Hi &   2351 &     32 & 7 &  -3.20 & -0.174 &    2.67 &  1 &  4900 & 0.90 &   3321 &   6275 \\
001134.52$+$155137.4 &  752-52251-378 & 4.325 & 20.09 & $-27.12$ &    H &   1173 &      0 & 7 &  -6.92 & -0.433 &    3.09 &  1 &  3080 & 0.74 &  27425 &  28948 \\
001306.15$+$000431.9 &  388-51793-624 & 2.165 & 18.48 & $-27.12$ &   Hi &    399 &      0 & 7 &   0.91 &  0.069 &    1.63 &  1 &  1378 & 0.61 &  13012 &  13732 \\
001328.21$+$135827.9 &  752-52251-204 & 3.576 & 18.94 & $-27.83$ &    H &    301 &      0 & 7 &   1.16 &  0.097 &    1.39 &  1 &  1381 & 0.31 &   3029 &   3717 \\
001400.45$+$004255.4 &  389-51795-412 & 1.710 & 18.34 & $-26.71$ &   Hi &    494 &      0 & 6 &  -0.24 & -0.019 &    1.29 &  1 &  2261 & 0.40 &  26174 &  27307 \\
001408.22$-$085242.2 &  652-52138-363 & 1.745 & 18.40 & $-26.72$ &   Hi &   3288 &      0 & 0 &  -0.23 &  0.203 &    0.00 &  1 &  4141 & 1.00 &   3866 &   5954 \\
001438.28$-$010750.1 &  389-51795-211 & 1.816 & 18.48 & $-26.76$ &   Hi &   3045 &      8 & 4 &   0.46 &  0.174 &    2.19 &  2 &  3381 & 1.00 &   3810 &   6155 \\
001502.26$+$001212.4 &  389-51795-465 & 2.852 & 18.84 & $-27.38$ &    H &   2198 &    717 & 6 &  -2.37 & -0.144 &    2.32 &  1 &  4415 & 1.00 &   7607 &  10127 \\
001528.86$-$090332.8 &  652-52138-403 & 2.000 & 19.26 & $-26.20$ &   HL &   6056 &   4834 & 7 &   1.54 &  0.282 &    1.41 &  1 &  8617 & 1.00 &  10374 &  15034 \\
001528.86$-$090332.8 &  652-52138-403 & 2.000 & 19.26 & $-26.20$ &   Lo &   1181 &    189 & 1 &   2.15 &  0.394 &    1.55 &  2 &  3310 & 0.52 &   8994 &  11698 \\
\enddata
\tablenotetext{a}{The SDSS spectrum is designated by its
plate-mjd-fiber.}
\tablenotetext{b}{The BAL subtype.  
The code n denotes a relatively narrow trough,
LoF denotes a FeLoBAL with \ion{Mg}{2} absorption in its SDSS spectra;
HLF denotes a FeLoBAL with \ion{C}{4} absorption in its SDSS spectra;
Lo denotes a LoBAL detected through \ion{Mg}{2} absorption; 
HL denotes a HiBAL in which broad ($\geq$1000~\kms) low-ionization
absorption is also seen;
Hi denotes a HiBAL-only object, in which broad low-ionization
absorption is {\em not} seen even though \ion{Mg}{2} is within the
spectral coverage,
and H denotes a HiBAL in which the \ion{Mg}{2} region is not within
the spectral coverage or has a very low signal-to-noise ratio and so
whether or not the object is a LoBAL as well as a HiBAL is unknown.}  
\tablenotetext{c}{The best-fitting template number.  Template numbers
0-3 correspond to the less luminous bin and numbers 4-7 to the more
luminous bin; line width increases from 0 to 3 and from 4 to 7.}
\tablenotetext{d}{The number of identified broad absorption troughs.}
\end{deluxetable}

\clearpage
\begin{deluxetable}{lrrrrl}
\tablecolumns{6}
\tablecaption{Manually Identified $z>4.38$ BALQSOs from the SDSS DR3 Quasar Catalog\label{tbl:hiz}}
\tablehead{
  \colhead{Quasar (SDSS J)} &
  \colhead{Spectrum\tablenotemark{a}} &
  \colhead{$z$} &
  \colhead{$i$} &
  \colhead{$M_i$} &
  \colhead{Type\tablenotemark{b}}
}
\startdata
001714.67$-$100055.4 &  652-52138-152 & 5.0105 & 19.523 & $-$27.935 &  H \\
012004.82$+$141108.2 &  424-51893-270 & 4.7293 & 20.199 & $-$27.131 &  H \\
082234.47$+$361534.8 &  826-52295-540 & 4.5995 & 19.934 & $-$27.385 &  H? \\
085634.92$+$525206.2 &  449-51900-246 & 4.8166 & 20.287 & $-$27.045 &  H \\
092819.28$+$534024.2 &  554-52000-620 & 4.3900 & 19.530 & $-$27.598 &  H \\
093333.86$+$051839.9 &  992-52644-156 & 4.4900 & 19.360 & $-$27.869 &  H \\
095151.18$+$594556.1 &  453-51915-586 & 4.8593 & 19.817 & $-$27.520 &  H? \\
101053.52$+$644832.0 &  488-51914-474 & 4.6848 & 19.848 & $-$27.469 &  H \\
102343.13$+$553132.3 &  946-52407-027 & 4.4550 & 19.315 & $-$27.835 &  HL \\
103601.03$+$500831.8 &  875-52354-474 & 4.4695 & 19.218 & $-$27.943 &  H \\
110041.96$+$580001.0 &  949-52427-031 & 4.7706 & 19.970 & $-$27.319 &  H? \\
110628.67$+$554054.8 &  908-52373-189 & 4.5481 & 20.056 & $-$27.136 &  H \\
110826.31$+$003706.7 &  278-51900-435 & 4.3950 & 19.780 & $-$27.385 &  H \\
111523.24$+$082918.4 & 1222-52763-241 & 4.6400 & 19.574 & $-$27.730 &  H \\
112956.09$-$014212.4 &  327-52294-337 & 4.8838 & 19.685 & $-$27.716 &  H \\
114448.54$+$055709.8 &  839-52373-413 & 4.7904 & 20.189 & $-$27.133 &  H? \\
120131.56$+$053510.1 &  842-52376-323 & 4.8303 & 19.455 & $-$27.875 &  H \\
120715.45$+$595342.9 &  954-52405-178 & 4.4833 & 20.157 & $-$27.026 &  H \\
121422.02$+$665707.5 &  493-51957-103 & 4.6389 & 18.876 & $-$28.370 &  H \\
123937.17$+$674020.8 &  494-51915-542 & 4.4235 & 20.170 & $-$26.975 &  H \\
124400.05$+$553406.8 & 1038-52673-371 & 4.6250 & 19.609 & $-$27.629 &  H \\
125850.93$+$615738.5 &  783-52325-316 & 4.4905 & 20.069 & $-$27.113 &  HL? \\
132853.66$-$022441.6 &  911-52426-266 & 4.6945 & 19.841 & $-$27.485 &  H \\
133304.50$+$604736.1 &  785-52339-149 & 4.4219 & 20.130 & $-$27.012 &  H? \\
135249.81$-$031354.3 &  914-52721-205 & 4.7477 & 19.798 & $-$27.555 &  H \\
141914.18$-$015012.6 &  917-52400-489 & 4.5860 & 19.092 & $-$28.215 &  H \\
142535.97$-$023934.4 &  918-52404-260 & 4.7500 & 19.425 & $-$27.975 &  H \\
142935.55$+$435628.9 & 1288-52731-248 & 4.6410 & 20.048 & $-$27.194 &  H \\
143352.20$+$022713.9 &  536-52024-346 & 4.7215 & 18.331 & $-$29.003 &  H? \\
144717.97$+$040112.4 &  587-52026-556 & 4.5800 & 19.168 & $-$28.091 &  H \\
151035.29$+$514841.0 & 1165-52703-475 & 5.0314 & 20.029 & $-$27.400 &  H \\
151404.78$+$473815.8 & 1330-52822-129 & 4.6673 & 19.702 & $-$27.589 &  H \\
160120.09$+$374948.9 & 1055-52761-308 & 4.4187 & 19.953 & $-$27.193 &  H \\
160501.20$-$011220.7 &  344-51693-054 & 4.9218 & 19.798 & $-$27.956 &  H \\
165354.61$+$405402.1 &  631-52079-127 & 4.9768 & 18.648 & $-$28.762 &  H \\
223521.22$-$082127.2 &  722-52224-388 & 4.4247 & 19.839 & $-$27.382 &  H \\
224147.75$+$135202.7 &  739-52520-557 & 4.4480 & 18.786 & $-$28.451 &  H\tablenotemark{c} \\
224255.52$+$124225.6 &  739-52520-059 & 4.4332 & 19.305 & $-$27.914 &  H \\
235152.80$+$160048.9 &  749-52226-417 & 4.6939 & 19.682 & $-$27.635 &  H? \\
\enddata
\tablenotetext{a}{The SDSS spectrum is designated by its
plate-mjd-fiber.}
\tablenotetext{b}{The Type column gives information on the subtype of
the BALQSO.  `H' denotes a HiBAL, `H?' denotes a candidate HiBAL, `HL'
denotes a confirmed LoBAL, and `HL?' a possible LoBAL.  Note that `H'
objects could still turn out to be LoBALs: a definitive classification
requires spectral coverage of the \ion{Al}{3} and \ion{Mg}{2} regions,
which is not available.}
\tablenotetext{c}{The \ion{C}{4} absorption trough in this object is
at $v>29,000$ km\,s$^{-1}$, and so this BALQSO would not have been
included in our main catalog even if it had been at $z<4.38$.}
\end{deluxetable}

\clearpage
%\newpage

%\begin{figure}
%\plotfiddle{histograms/deepestVi.ps}{0pt}{0}{470}{380}{0}{0}
%\caption{Plots of the deepest absorption with absolute i magnitude in
%BALQSOs. Few troughs of shallow depth (0.1-0.3) are identified at
%less luminous absolute magnitudes where the signal-to-noise ratio is
%lower. \label{fig:deepestVi}}
%\end{figure}

\begin{thebibliography}{}

\bibitem[Abazajian et~al. 2003]{sdssdr1}
  Abazajian, K., et~al. 2003, AJ, 126, 2081

\bibitem[Abazajian et~al. 2005]{sdssdr3}
  Abazajian, K., et~al. 2005, AJ, 129, 1755

\bibitem[Baldwin 1977]{baldwin}
  Baldwin, J.~A. 1977, ApJ, 214, 679

\bibitem[Becker et~al. 2000]{cocoon}
  Becker, R.~H., White, R.~L., Gregg, M.~D., Brotherton, M.~S. et
  al. 2000, ApJ, 538, 72

\bibitem[Blanton et~al. 2003]{TilingAlgorithm}
  Blanton, M.~R., Lupton, R.~H., Maley, F.~M., Young, N., Zehavi, I.,
  and Loveday, J. 2003, AJ, 125, 2276

\bibitem[Boroson \& Meyers 1992]{cocoon2}
  Boroson, T.~A. \& Meyers, K.~A. 1992, ApJ, 397, 442

\bibitem[Brandt, Laor, \& Wills 2000]{xraybals}
  Brandt, W.~N., Laor, A., Wills, B.~J. 2000, ApJ 528, 637

\bibitem[Cardelli, Clayton, \& Mathis 1989]{deredden}
  Cardelli, J.~A., Clayton, G.~C., and Mathis J.~S. 1989, ApJ 345, 245

\bibitem[Churchill et~al. 2000]{churchill}
  Churchill, C.~W., Mellon, R.~R., Charlton, J.~C., Jannuzi, B.~T.,
  Kirhakos, S., Steidel, C.~C., \& Schneider, D.~P. 2000, ApJ, 543,
  577

\bibitem[Croom et~al. 2004]{qsolumfunc}
  Croom, S.~M., Smith, R.~J., Boyle, B.~J., Shanks, T., Miller, L.,
  Outram, P.~J., \& Loaring, N.~S. 2004, MNRAS, 349, 1397

\bibitem[Ding et~al. 2003]{ding}
  Ding, J., Charlton, J.~C., Churchill, C.~W., \& Palma, C. 2003, ApJ,
  590, 746

%\bibitem[Foltz et~al. 1986]{foltz}
%  Foltz, C.~B., Weymann, R.~J., Peterson, B.~M., Sun, L., Malkan,
%  M.~A., \& Chaffee, F.~H., Jr. 1986, ApJ, 307, 504

\bibitem[Fukugita et~al. 1996]{ugriz}
  Fukugita, M., Ichikawa, T., Gunn, J.~E., Doi, M., Shimasaku, K., \&
  Schneider, D.~P. 1996, AJ, 111, 1748

\bibitem[Gallagher et~al. 2002]{balxray2}
  Gallagher, S.~C., Brandt, W.~N., Chartas, G., \& Garmire,
  G.~P. 2002, ApJ, 567, 37

\bibitem[Ganguly et~al. 2001]{nals}
  Ganguly, R., Bond, N.~A., Charlton, J.~C., Eracleous, M., Brandt,
  W.~N., \& Churchill, C.~W. 2001, ApJ, 549, 133

\bibitem[Gunn et~al. 1998]{camera}
  Gunn, J.~E., et~al. 1998, AJ, 116, 3040

\bibitem[Gunn et~al. 2006]{telescope}
  Gunn, J.~E., et~al. 2006, AJ, 131, in press

\bibitem[Hall et~al. 2002]{balai}
  Hall, P.~B., Anderson, S.~F., Strauss, M.~A., York,
  D.~G. et~al. 2002, ApJS, 141, 267

\bibitem[Hall et~al. 2006]{analysis}
  Hall, P.~B., Trump, J.~R., Richards, G.~T., Proga, D., Reichard,
  T.~A., Vanden Berk, D.~E., \& Schneider, D.~P. 2006, in prep.

\bibitem[Hazard et~al. 1984]{hazard}
  Hazard, C., Morton, D.~C., Terlevich, R., \& McMahon, R. 1984, ApJ,
  282, 33

%\bibitem[Hewett, Foltz, \& Chaffee 1995]{LBQS}
%  Hewett, P.~C., Foltz, C.~B., \& Chaffee, F.~H. 1995, AJ, 109, 1498

\bibitem[Hewett \& Foltz 2003]{LBQSbals}
  Hewett, P.~C. \& Foltz, C.~B. 2003, AJ, 125, 1784

\bibitem[Hogg et~al. 2001]{hogg}
  Hogg, D.~W., Schlegel, D.~J., Finkbeiner, D.~P., \& Gunn, J.~E. 2001,
  AJ, 122, 2129

\bibitem[Hopkins et~al. 2004]{hopkins}
  Hopkins et~al. 2004, AJ, 128, 1112

\bibitem[Ivezi\'c et~al. 2004]{QA}
  Ivezi\'c, Z., et~al. 2004, AN, 325, 583 

\bibitem[Lee \& Turnshek 1995]{lee_turn}
  Lee, L.~W. \& Turnshek, D.~A. 1995, ApJL, 453, L61

\bibitem[Lupton, Gunn, \& Szalay 1999]{asinh}
  Lupton, R.~H., Gunn, J.~E., \& Szalay, A.~S. 1999, 1999, AJ, 118,
  1406

\bibitem[Lupton et~al. 2001]{lupton}
  Lupton, R.~H., Gunn, J.~E., Ivezi\'c, \v{Z}., Knapp, G.~R., Kent, S.,
  \& Yasuda, N. 2001, in ASP Conf. Ser. 238, Astronomical Data
  Analysis Software and Systems, eds. F.~R. Harnden, F.~A. Primini, \&
  H.~E. Payne (San Francisco:ASP), 269

\bibitem[Marziani et~al. 2003]{marziani-sulentic}
  Marziani, P., Sulentic, J.~W., Zamanov, R., Calvani, M.,
  Dultzin-Hacyan, D., Bachev, R., \& Zwitter, T. 2003, ApJS, 145, 199

\bibitem[Menou et~al. 2001]{menou}
  Menou, K., et~al. 2001, ApJ, 561, 645

\bibitem[Murray \& Chiang 1998]{AGNmodel}
  Murray, N. \& Chiang, J. 1998, ApJ, 494, 125

\bibitem[Osmer, Porter, \& Green 1994]{osmer}
  Osmer, P.~S., Porter, A.~C., \& Green, R.~F. 1994, ApJ, 436, 678

\bibitem[Pei 1992]{peismc}
  Pei, Y.~C. 1992, ApJ, 395, 130

\bibitem[Peterson 2003]{AGNmodel2}
  Peterson, B.~M. 2003, in ASP Conf. Ser. 290, Active Galactic Nuclei:
  From Central Engine to Host Galaxy, eds. S. Collin, F. Combes \&
  I. Shlosman (San Francisco:ASP), 43

\bibitem[Pier et~al. 2003]{pier}
  Pier, J.~R., Munn, J.~A., Hindsley, R.~B., Hennessy, G.~S., Kent,
  S.~M., Lupton, R.~H., \& Ivezi\'c, \v{Z}., 2003, AJ, 125, 1559

\bibitem[Press et~al. 1992]{numrecipes}
  Press, W.~H., Teukolsky, S.~A., Vetterling, W.~T., \& Flannery,
  B.~P.  1992, Numerical recipes in C. The art of scientific computing
  (Cambridge: University Press, 1992, 2nd ed.)

\bibitem[Prevot et~al. 1984]{prevotsmc}
  Prevot, M.~L., Lequeux, J., Prevot, L., Maurice, E., \&
  Rocca-Volmerange, B. 1984, A\&A, 132, 389

\bibitem[Reichard et~al. 2003]{baledr}
  Reichard, T.~A., et~al. 2003, AJ, 120, 1711

\bibitem[Reichard et~al. 2003b]{baledr2}
  Reichard, T.~A., et~al. 2003b, AJ, 126, 2594

\bibitem[Richards et~al. 2001]{gtr01}
  Richards, G.~T., et~al. 2001, AJ, 121, 2308

\bibitem[Richards et~al. 2002]{gtr02}
  Richards, G.~T., et~al. 2002, AJ, 123, 2945

\bibitem[Richards et~al. 2003]{gtr03}
  Richards, G.~T., Hall, P.~B., Reichard, T.~A., \& Vanden Berk,
  D.~E. 2004, in ASP Conf. Ser. 311, AGN Physics with the SDSS
  Conf. Proc., eds. G.~T. Richards \& P.~B. Hall, (San Francisco:ASP),
  25

\bibitem[Richstone \& Schmidt 1980]{qsopowerlaw}
  Richstone, D.~O. \& Schmidt, M. 1980, ApJ, 235, 361

\bibitem[Schlegel, Finkbeiner, \& Davis 1998]{reddening}
  Schlegel, D.~J., Finkbeiner, D.~P., \& Davis, M. 1998, ApJ, 500, 525

\bibitem[Schneider et~al. 2002]{edrqso}
  Schneider, D.~P., et. al. 2002, AJ, 123, 567

\bibitem[Schneider et~al. 2003]{dr1qso}
  Schneider, D.~P., et. al. 2003, AJ, 126, 2579

\bibitem[Schneider et~al. 2005]{dr3qso}
  Schneider, D.~P., et. al. 2005, AJ, 130, 367 %(astro-ph/0503679)

\bibitem[Sigut, Pradhan, \& Nahar 2004]{sigut}
  Sigut, T.~A.~A., Pradhan, A.~K., \& Nahar, S.~N. 2004, ApJ, 661, 81

\bibitem[Smith et~al. 2002]{smith}
  Smith, J.~A., et~al. 2002, AJ, 123, 2121

\bibitem[Sprayberry \& Foltz 1992]{2175bump}
  Sprayberry, D. \& Foltz, C.~B. 1992, ApJ, 390, 39

\bibitem[Spergel et~al. 2003]{wmap}
  Spergel, D.~N., et~al. 2003, ApJ, 148, 175

\bibitem[Stoughton et~al. 2002a]{edr}
  Stoughton, C. et~al. 2002a, AJ, 123, 485

\bibitem[Stoughton et~al. 2002b]{dataquery}
  Stoughton, C. et~al. 2002b, SPIE, 4836, 339

\bibitem[Tolea, Krolik \& Tsvetanov 2002]{tkt}
  Tolea, A., Krolik, J.~H., \& Tsvetanov, Z. 2002, ApJL, 578, 31

\bibitem[Tucker et~al. 2006]{MT Pipeline}
  Tucker, D., et~al. 2006, PASP, submitted 

\bibitem[Vanden Berk et~al. 2001]{edrcomp}
  Vanden Berk, D.~E. et~al. 2001, AJ 122, 549

\bibitem[Vanden Berk et~al. 2003]{dr1spec}
  Vanden Berk, D.~E., Yip, C.~W., Connolly, A.~J., Jester, S., \&
  Stoughton, C. 1994, AGN Physics with the SDSS Conf. Proc.,
  eds. G.~T. Richards \& P.~B. Hall, 21

\bibitem[Vestergaard \& Wilkes 2001]{MgIInorm}
  Vestergaard, M. \& Wilkes, B.~J. 2001, ApJS, 134, 1

\bibitem[Voit, Weymann, \& Korista 1993]{mgiitrough}
  Voit, G.~M., Weymann, R.~J., \& Korista, K.~T. 1993, ApJ, 413, 95

\bibitem[Weymann et~al. 1991]{BI}
  Weymann, R.~J., Morris, S.~L., Foltz, C.~B., \& Hewitt, P.~C. 1991,
  ApJ, 373,23

\bibitem[York et~al. 2000]{sdssrev}
  York, D.~G., et~al. 2000, AJ, 120, 1579

\end{thebibliography}
\end{document}